\documentclass[useAMS,usenatbib]{mn2e}
\usepackage{amsmath}
\usepackage{txfonts}
\usepackage{graphicx}
\usepackage{times}
\usepackage{color}

\newcommand{\apj}{ApJ}

\newcommand{\mnras}{MNRAS}

\newcommand{\physrep}{Physics Reports}

\begin{document}
\voffset-1.25cm
\title[Quick Particle Mesh]{Mock galaxy catalogs using the quick
particle mesh method}
\author[White et al.]{
\parbox{\textwidth}{Martin White$^{1,2}$, Jeremy L.~Tinker${}^3$ and
Cameron K.~McBride${}^4$}
\vspace*{4pt} \\
$^{1}$ Lawrence Berkeley National Laboratory, 1 Cyclotron Road,
Berkeley, CA 94720, USA \\
$^{2}$ Departments of Physics and Astronomy, University of California,
Berkeley, CA 94720, USA \\
$^{3}$ Center for Cosmology and Particle Physics, Department of Physics,
New York University, New York, NY 10003, USA \\
$^{4}$ Harvard-Smithsonian Center for Astrophysics, 60 Garden Street,
Cambridge, MA 02138, USA
}

\date{\today} 
\pagerange{\pageref{firstpage}--\pageref{firstpage}}

\maketitle

\label{firstpage}

\begin{abstract}
Sophisticated analysis of modern large-scale structure surveys requires
mock catalogs.  Mock catalogs are used to optimize survey design, test
reduction and analysis pipelines, make theoretical predictions for basic
observables and propagate errors through complex analysis chains.  We
present a new method, which we call ``quick particle mesh'', for generating
many large-volume, approximate mock catalogs at low computational cost.
The method is based on using rapid, low-resolution particle mesh simulations
that accurately reproduce the large-scale dark matter density field.
Particles are sampled from the density field based on their local density such
that they have $N$-point statistics nearly equivalent to the halos resolved
in high-resolution simulations, creating a set of mock halos that can be
populated using halo occupation methods to create galaxy mocks for a variety
of possible target classes.
\end{abstract}


\pagebreak

\section{Introduction}
\label{sec:intro}

The study of the large-scale structure in the Universe is a cornerstone
of modern cosmology. In addition to allowing us to understand the structure
itself, such studies offer an incisive tool for probing cosmology and particle
physics and set the context for our modern understanding of galaxy formation
and evolution.  Large galaxy surveys play a key role in this enterprise, and
ever larger surveys have provided increasing insight and ever tighter
constraints on cosmological models.
For almost as long as there have been sky surveys, people have used mock
catalogs in order to interpret them
\citep[e.g.][]{NeyScoSha53,ScoShaSwa54,SonPee77,SonPee78,Sha79}.
With time, surveys have become larger and the type of questions we ask from
them have changed significantly, becoming steadily more quantitative.
Driven by the need for ever more precise theoretical predictions and the use
of increasingly complex and sophisticated data-analysis algorithms,
simulations and synthetic data sets have become increasingly important.

Within the modern paradigm, wherein galaxies and quasars form and evolve
within the halos of the cosmic web, the natural technique for creating
mock catalogs is N-body simulation
\citep[see e.g.][for a recent review]{LargeScaleStructure}.
N-body simulations have been the workhorse of modern cosmology for several
decades, and codes with high force resolution can accurately predict the
abundance, spatial distribution, profiles and substructure of dark matter
halos in representative cosmological volumes \citep[e.g.][]{Millennium,Kuh12}.
Such runs are often quite expensive, however, and while it is certainly
possible \citep[e.g.][]{MICE,LasDamas,Coyote} the practicality of running
very large numbers of them for Monte-Carlo studies is unclear\footnote{Or
at the very least, it limits the applicability to a small number of groups.}.
This is especially true if the majority of the information can be obtained
more cheaply.
In this vein there are two important considerations: the total run-time of
any given method and the total memory requirement.
The latter is in many ways the most stringent if we wish to enable the
generation of many simulations on commonly available hardware.

Over the years a large number of approximate methods have been developed
which produce halo catalogs of sufficient reliability for many tasks, such
as predicting the very large-scale distribution of galaxies in simulations.
Many of these methods started out as a means to highlight key
characteristics of non-linear structure formation and were implemented
numerically (this includes such methods as the adhesion approximation
[\citealt{GurSaiSha89,WeiGun90,KPSM92,MelShaWei94,ValBer11}];
the log-normal model [\citealt{ColJon91}];
the truncated Zel'dovich approximation [\citealt{ColMelSha93,MelPelSha94}];
the frozen flow approximation [\citealt{Mat92}];
the free-particle approximation [\citealt{ShoCol06}];
PThalos [\citealt{PThalos,Man13}];
Pinocchio [\citealt{MonTheTaf02,Mon13}]);
remapping Lagrangian perturbation theory [\citealt{Lec13}]; 
PATCHY [\citealt{PATCHY}]; 
and machine-learning techniques [\citealt{Xu13}]).
A recent summary of some of these methods can be found in \citet{Ney12}.
One of the key observations is that much of the filamentary cosmic web of
structures which arises within N-body simulations is also present in these
more approximate methods.  Indeed, much of the ``work'' involved in running
an N-body simulation in a large volume is evolving linear theory, or modes
which are not far from linear.
This gives some hope that approximate and cheap algorithms can capture many
of the important properties of full N-body simulations.

The most widely used approximation is the log-normal model, which has
been used to create mock catalogs for many large-scale structure
surveys in the past decade \citep[e.g.,][among
others]{Col05,Per10,Rei10,Bla11,Beu11,Beu12}.  By contrast, the BOSS
team used mocks based on second-order Lagrange perturbation theory
in their recent cosmology analyses (PThalos; \citealt{Man13}).

In this paper we present a new, approximate, method for generating
mock galaxy catalogs for the first step in a Monte-Carlo simulation of
redshift surveys.
There are two particular arguments which pushed us to consider
the rapid generation of mock catalogs as a means to Monte-Carlo
our errors. 
The first is that many steps in the analysis can be quite non-linear,
for example the reconstruction procedure for BAO which involves
constrained realizations of interpolating fields and non-linear motions
of both galaxy and random points prior to computing the 2-point
function.  The interaction of these non-linearities with the complex
observing geometry can be particularly difficult to model directly.
The second is that we are primarily interested in quite large scales,
where the complexities of galaxy formation and fully non-linear clustering
are mitigated.  On smaller scales internal estimates of the covariance
matrix can be constructed (e.g.~via bootstrap or jackknife), however on
large scales these methods do not perform as well
\citep[e.g.][and references therein]{Nor09}.
For these reasons we wish to investigate a procedure or set of procedures
which allow us to generate point sets which mimic the observational samples,
at least in terms of low order statistical properties.  As is often the
case, increasing fidelity for each simulation comes at a price of increased
computational complexity which implies fewer realizations (and more noise
in the Monte-Carlo, see e.g.~\citealt{TJK13}) for a fixed computational effort.

The outline of this paper is as follows.  
In Section~\ref{sec:mock}, we discuss different approaches to producing
mock catalogs and the direct N-body simulation method which shall serve
as our benchmark.  
Section~\ref{sec:qpm} details the ``quick particle mesh'' (QPM) method, 
the main substance of this paper. 
Section~\ref{sec:comparison} compares the mock catalogs produced by a
variety of different methods, including QPM, to those derived from an
N-body simulation.  
We briefly describe the public implementation of this method, 
\texttt{mockFactory}, in Section~\ref{sec:mockFactory}.
Finally, we discuss limitations, extensions, and comparisons to other 
methods recently presented in Section~\ref{sec:summary}. 
Unless otherwise stated, the assumed cosmology is flat $\Lambda$CDM with 
$\Omega_m=0.274$, $\Omega_b=0.046$, $\Omega_\Lambda=0.726$, $h=0.7$, $n=0.95$, 
and $\sigma_8=0.8$.  

\section{Mock catalog methods} 
\label{sec:mock}

Modern galaxy formation models assume that galaxies form and remain in
the potential wells of dark matter halos.
The four basic steps of creating mock galaxy catalogs are
(1) predicting the evolution of the underlying mass field,
(2) locating and characterizing the properties of dark matter halos,
(3) populating the halos with mock galaxies and
(4) applying survey characteristics to the box of galaxies.
The halo to mock galaxy mapping is often trained by matching small-scale
clustering statistics to the observational galaxy sample
\citep[e.g.,][]{ReiSpe09,Whi11}.
The last step involves adding survey-specific realism to the mock
galaxy distribution, such as applying radial and angular selection functions
to match the geometry.
The final goal is to produce a set of points that statistically matches
the spatial distribution of observational galaxy samples, and in our case
to be able to do this many times so as to characterize the probability
distributions of observables.

Characterizing dark matter halos, while not strictly required, is a useful
step. By definition, halos represent overdense regions of the mass field
(typically 100-300 times the mean density or 200 times the critical density)
that arise from non-linear gravitational collapse.
Approximate methods of predicting the mass evolution do not attempt to
accurately calculate this collapse.
Instead the focus is on larger scales dominated by linear and weakly
non-linear gravitational evolution.  The strongly non-linear 
interactions within halos, which are represented in the observational galaxy 
distributions, can be added with analytic prescriptions on top of the halo 
distribution.  If adequate halo catalogs can be generated, the subsequent steps 
towards survey-specific mock galaxy catalogs are comparatively straightforward. 

For this work, we will compare various fast methods of halo catalog generation
against the relevant properties of halo catalogs derived from periodic box
$N$-body simulations.  We have implemented three approximate methods of mock
catalog creation.  Two of these represent recent methods used to model galaxy
surveys, namely the log-normal (LGN) model and the method of \citet{Man13},
based on second order Lagrangian Perturbation theory, which we shall refer to
as LPT.  We discuss the details of our implementations in
Appendix~\ref{app:mocks}.  We compare both of these models to a our
new method which we refer to as ``quick particle mesh'' or QPM which
we describe in detail in Section~\ref{sec:qpm}.

The focus of our investigation resolves halo masses and volumes appropriate to 
current modern redshift surveys, specifically galaxies in SDSS-III BOSS (Dawson 
et al 2012).  This is a practical choice; the application of these methods is 
certainly not restricted to modeling these types of galaxies. Indeed,
our method is flexible enough to create mocks for a variety of survey
specifics and target types, even within the same mock galaxy distribution.

\subsection{N-body simulations}
\label{sec:Nbody}

We make use of several N-body simulations in this paper,  each of the
$\Lambda$CDM family with the same cosmology ($\Omega_m=0.274$,$\Omega_b=0.046$, 
$\Omega_\Lambda=0.726$, $h=0.7$, $n=0.95$, and $\sigma_8=0.8$).
These (high resolution) simulations will form our benchmark and be the fiducial 
model of ``truth''.  The have also been used in \citet{Whi11,ReiWhi11,Whi12} and more 
details can be found in those papers.

Our high resolution simulations resolve all of the halos described
throughout the paper.  Briefly, we used an updated version of the
TreePM\footnote{This TreePM code has been compared to a number of other
codes and shown to perform well for such simulations \citep{Hei08}.
The code has been modified to use a hybrid MPI+OpenMP approach which is
particularly efficient for modern computing platforms.}
code described in \citet{TreePM} to evolve $3000^3$ particles
($5.9\times 10^{10}\,h^{-1}M_\odot$) in a box of side
$2750\,h^{-1}$Mpc.  We ran 2 realizations of this simulation,
differing only in the random number seed chosen for the initial
conditions.  We also ran 20 realizations of the same cosmology using
$1500^3$ equal mass ($7.6\times 10^{10}\,h^{-1}M_\odot$) particles in
a periodic cube of side length $1500\,h^{-1}$Mpc.  This second set
allow for better estimation of the sample variance at large scales.
For each simulation, the initial conditions were generated by displacing
particles from a regular grid using second order Lagrangian perturbation
theory \citep{Buc89,Mou91,Hiv95} at $z=75$, where the rms displacement is
$10$ per cent of the mean inter-particle spacing.

For each output we found dark matter halos using a friends-of-friends 
algorithm \citep[FoF; ][]{DEFW}
with a linking length of $0.168$ times the mean interparticle spacing.
This partitions the particles into equivalence classes roughly bounded by
isodensity contours of $100\times$ the mean density.
The position of the most-bound particle and the center of mass velocity are
stored for each halo and used in the comparisons described below.

\section{QPM: the method} 
\label{sec:qpm}

The QPM method uses a low resolution particle-mesh (PM) N-body solver to
evolve particles within a periodic simulation volume as is common in the field.
The time steps are set to be quite large and the mesh scale and mean
inter-particle spacing exceed the size of all but the largest dark matter 
halos.
In this manner we keep both the run time and the memory requirements modest.

The particle-mesh evolution of $N$ particles within a periodic cube
employs fast Fourier transforms on a fixed Cartesian mesh to compute
the force.  In our configuration, the force mesh has as many nodes as
the simulation has particles, and we use $2\,h^{-1}$Mpc as the mean
inter-particle spacing as our default.  We start each simulation at
$z=25$ using second order Lagrangian perturbation theory.
The rms particle displacement is 15 per cent of the mean
inter-particle spacing at this redshift.  The code evolves the
particles using a second order leap-frog method, with time steps of
$\Delta\ln a=15$ per cent.  At each step the potential on the grid is
computed using a $1/k^2$ kernel and the force is derived from the
potential using $4^{\rm th}$ order differencing.

We experimented with different choices for the time step, number of particles
and mesh spacing.  The choice of time step is a major driver of run time and
the choice of mesh spacing is a major factor in memory requirements.
Since we are primarily interested in quite large scales we found that even
very large time steps and coarse force meshes were sufficient to get
convergence to the required accuracy in our mock halo catalogs.
Setting the mesh and particle numbers equal was convenient.
Our choice of 15 per cent in $\ln a$ provided a good trade-off between
adequately resolving the change in the growth of structure during the
matter-dominated to $\Lambda$-dominated transition and speed.
Steps in $a$ and in the linear growth rate produced similar behavior,
we decided on constant steps in $\ln a$ partly out of simplicity.


In QPM we only resolve the density field on large-scales and so must
decide how to partition the mass, which in hierarchical theories forms
in a bottom-up manner, into a spectrum of dark matter halos using only
large-scale information.  We have chosen to do this using the local
density for each particle at the time of relevance\footnote{We are not
restricted to using only the local density smoothed on a single scale in
deciding where to place halos in our QPM field.  We can use the density
smoothed on multiple scales, the density field in the initial conditions, or
other quantities (e.g.~the tidal shear or the initial peak curvature) as well.
We have not explored this issue further, but it is an interesting question
which set of large-scale observables of the matter field best predicts the
existence of a halo in a higher resolution simulation.}.  Based on its
density we annoint a subset of the simulation particles as mock dark
matter halos, recording the position and velocity of these ``halos''.
Selecting particles based on their local density allows us to work
with low particle numbers and hence low computational cost as well as
offering a different avenue to explore compared to full simulation.
This is similar in spirit to the method used in \citet{CHWF98} or
\citet{Wec04}, although the technical details differ somewhat.

There is no reason, in principle, why we cannot skip the halo creation
step and generate a sample of mock ``galaxies'' directly by sampling
particles.  However we have found it very useful to have halo
information for our samples, and it allows tighter connection with
halo-occupation modeling, as well as the ability to model multiple
target samples within the same mock---e.g., red and blue galaxies, or
bright and faint galaxies---essentially, and target population that
can be modeled within the halo occupation context. Thus we shall
always go through halos in this paper.

To keep the run-time as small as possible the density is estimated using
Fourier methods on the same mesh as is used for the force calculation.
The density is interpolated onto and off of the mesh using clould-in-cell
interpolation, and the density is smoothed with a Gaussian kernel of 1 mesh
cell in width (i.e.~$\sigma=2\,h^{-1}$Mpc).
If the smoothing kernel is too large then a scale-dependent bias is
introduced\footnote{Sampling
  particles with probability $p(\delta_{\rm smth})$ is the same as
  computing the correlation function of
  $(1+\delta_{\rm part})p(\delta_{\rm smth})$, where $\delta_{\rm part}$
  is the density field defined by the particles (i.e.~unsmoothed).  In this
  correlation function mode-coupling introduces scale-dependent bias.}
near the acoustic peak in the correlation function at $100\,h^{-1}$Mpc.
For a $2\,h^{-1}$Mpc Gaussian smoothing the scale-dependence is small
enough that it is not a concern.
Once the density field is computed using all of the particles, the position,
velocity, and density of a random subset of the particles are saved.  


\begin{figure}
\begin{center}
\resizebox{3.25in}{!}{\includegraphics{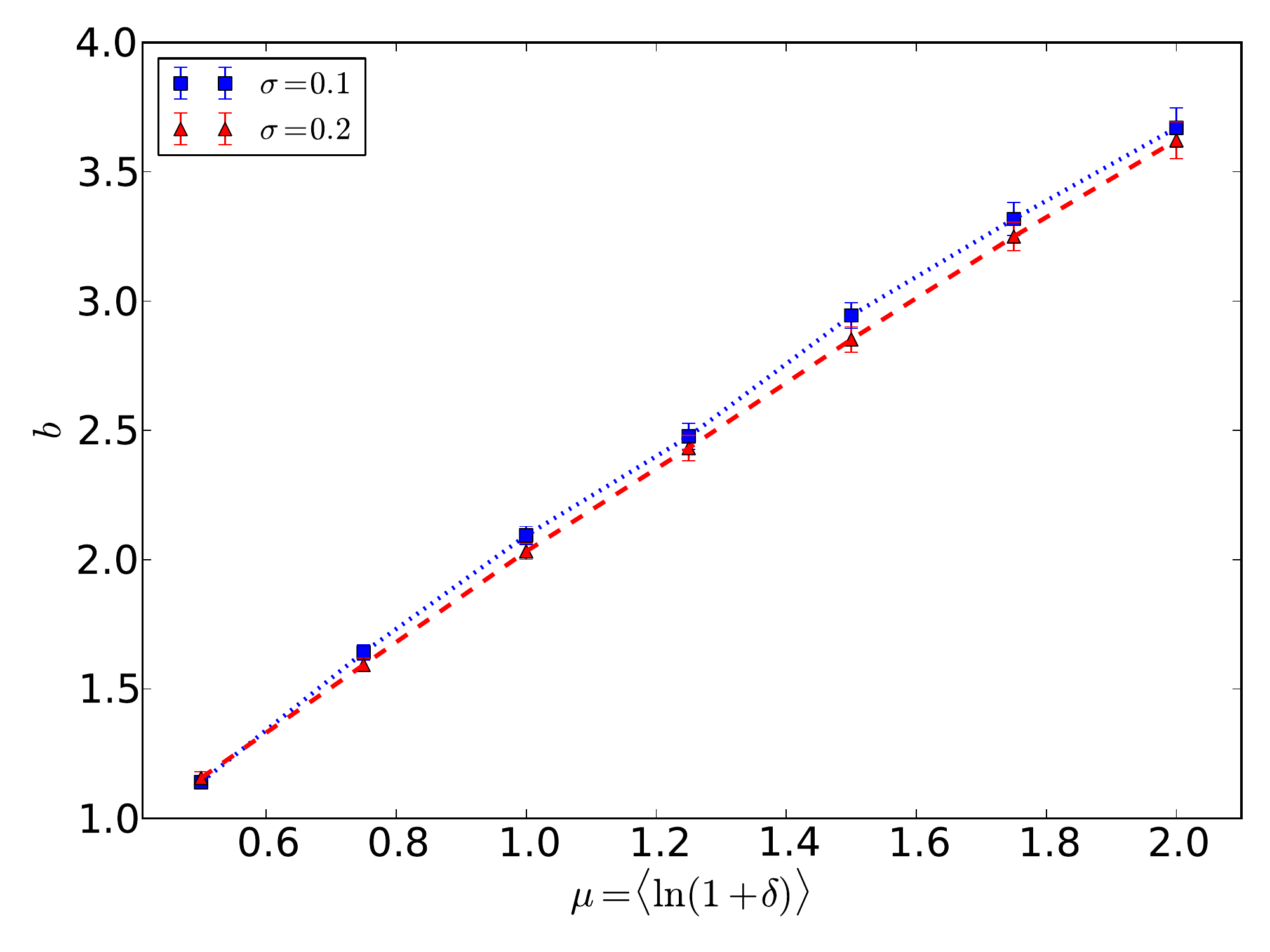}}
\end{center}
\caption{The large-scale bias, $b$, of particles selected based on their local
density.  The probability of selection is a Gaussian in $\ln(1+\delta)$
with mean $\mu$ and width $\sigma$.  We show the bias as a function of $\mu$
for two values of $\sigma$: $0.1$ (blue squares and dotted lines) and $0.2$
(red triangles and dashed lines).}
\label{fig:b_vs_mu}
\end{figure}

As we intend these particles to stand in for halos, it might be better to store
the average velocity smoothed on some scale rather than the particular
velocity of the particle.  However, at the scales of interest and the
resolutions we choose, we are relatively insensitive to ``virial'' motions
within halos, so we use the particle velocity for simplicity.
In fact, comparing the pairwise velocity dispersion of particles in the PM
simulations with those in a higher resolution simulation we see that the
small-scale velocity field is not well resolved by the PM simulation.
Thus we add an additional, Gaussian random velocity of
$125\,{\rm km}\,{\rm s}^{-1}$ to each component of the halo velocity.
As this value is dependent on the details of the simulation we make it a
free parameter in the model and the software described below.  It affects
the measured quadrupole moment of the correlation function on small scales
and can in principle be adjusted to improve agreement with measurements on
small scales, if desired.

\begin{figure}
\begin{center}
\resizebox{3.25in}{!}{\includegraphics{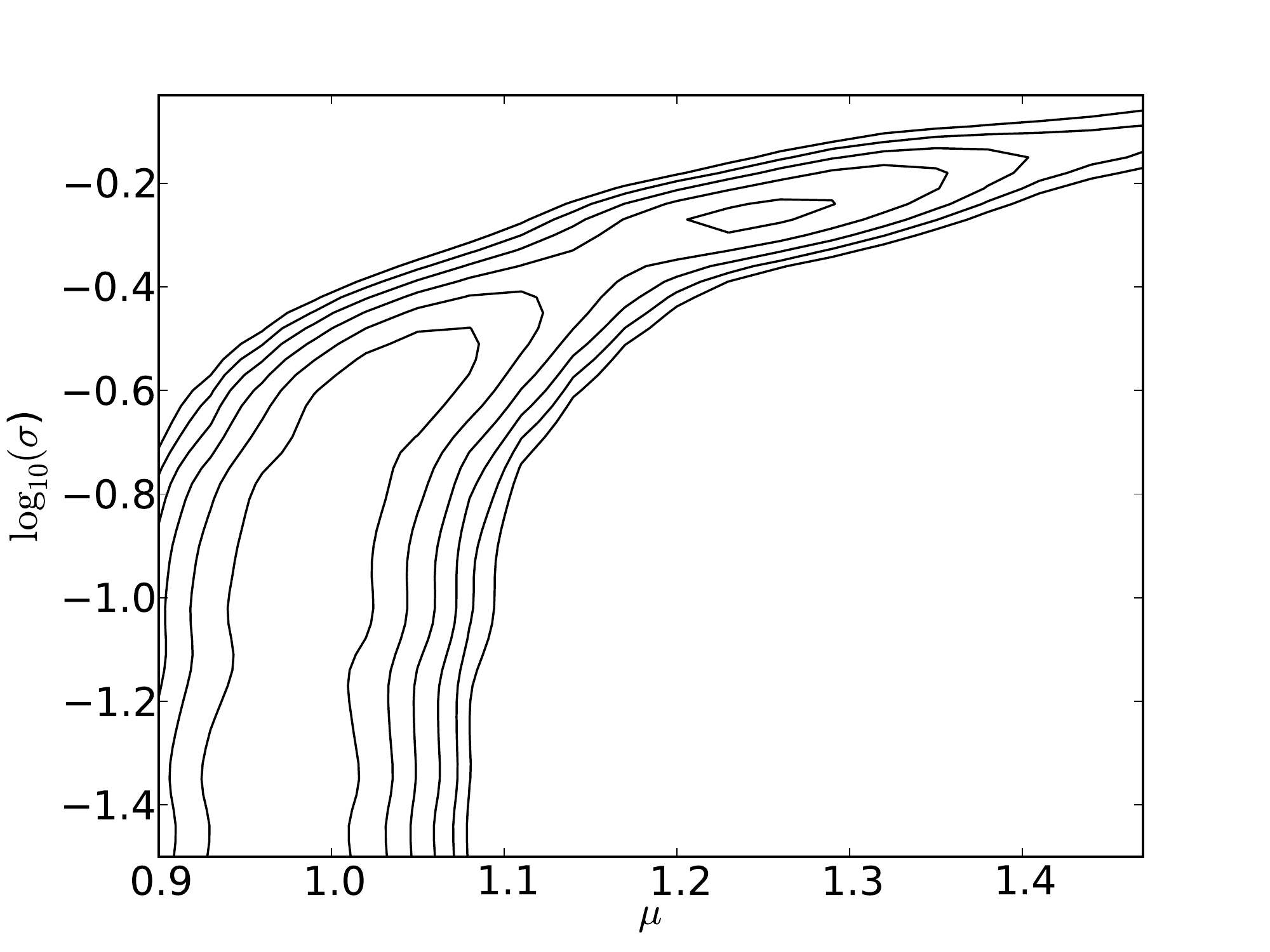}}
\end{center}
\caption{Contours of the likelihood for fitting the large-scale correlation
function as a function of scale $\mu$ and width $\sigma$.  Contours are spaced
by $\sqrt{2}$ in likelihood, from the peak.
In this example we matched the correlation function of the selected particles
to $2^2$ times the real-space, matter correlation function.  The fit was
over the range $30<r<60\,h^{-1}$Mpc with 10 bins and 10 per cent (uncorrelated)
errors per bin.}
\label{fig:lik_cont}
\end{figure}

The mock halo catalog at each output time is constructed in post-processing.
Our goal is to choose particles from the saved subset, with a density-dependent
probability, and have them stand in for halos of a given mass.
We select the particles and assign halo masses so as to match the mass
function and large-scale bias of halos, as determined from high-resolution
simulations \citep[e.g.][]{Tin08,Tin10}.
The choice of sampling function is arbitrary, and we tried several.
A convenient form is a Gaussian in $\ln(1+\delta)$, with a mean $\mu$ and
a width $\sigma$ which we can adjust to get the desired clustering.
Higher $\mu$ in general leads to a higher large-scale bias, as shown in
Fig.~\ref{fig:b_vs_mu}.

In our fiducial models we held $\sigma=0.1$ fixed and adjusted $\mu(b)$
so as to reproduce the large-scale bias, $b$, as a function of halo mass
(taken from \citealt{Tin10}).
The same large-scale clustering can be achieved by simultaneously increasing
$\mu$ and $\sigma$ (see Fig.~\ref{fig:lik_cont}), but we are more interested
in the models with lower $\sigma$ and holding $\sigma$ fixed is a convenient
choice.

We divided the range of halo masses we wish to produce into $N_h$ bins spaced
equally in $\log M_h$.
The results are insensitive to the value of $N_h$ as long as $b(M_h)$
doesn't vary significantly over the bin.
For each mass bin we know how many particles we need to sample to match the
mass function \citep{Tin08} and we know the Gaussian we need to sample from
to match the desired large-scale bias (from the calibration of $\mu(b)$ above).
Thus we simply loop over the particles and select halos in a mass bin from
them with the necessary probability.
The resulting set of mock halos has the proper number density and
large-scale bias as a function of mass.  Note that this procedure is
equivalent to using Bayes' theorem in the form
$P(M_h|x)\propto P(x|M_h)P(M_h)$ with $x=\ln(1+\delta)$.

It is of course possible to simultaneously adjust both $\mu$ and $\sigma$ as
a function of halo mass to better match the scale-dependence of halo bias at
smaller scales, but we did not find that to be useful for our purposes.


Recently \citet{COLA} introduced a rapid simulation method which they
referred to as ``COLA:  COmoving Lagrangian Acceleration''.
This method accelerates a PM code by solving for the evolution of large-scale
structure in a frame that is comoving with trajectories calculated in
Lagrangian Perturbation Theory (LPT).
While our procedure is quite similar in spirit, we have chosen not to
implement this speed-up for three reasons.
The first is that the run-time of our simulations is already short (taking
only a few more steps than COLA).
The second is that the COLA method carries with it memory overhead
(holding the first and second order Lagrangian displacements for each particle)
and memory is the primary driver for simulation size or volume in our situation.
Finally the COLA method, like any PM scheme, resolves halos and halo positions
reliably only when the mesh scale is significantly finer than the mean
inter-particle separation and the halo virial radius.  If run in this regime
over very large volumes the code requires a large amount of memory to hold the
force mesh, and this severely limits the machines upon which it can be run.

\section{Comparison of methods} \label{sec:comparison}

\begin{figure*}
\begin{center}
\resizebox{7.5in}{!}{\includegraphics{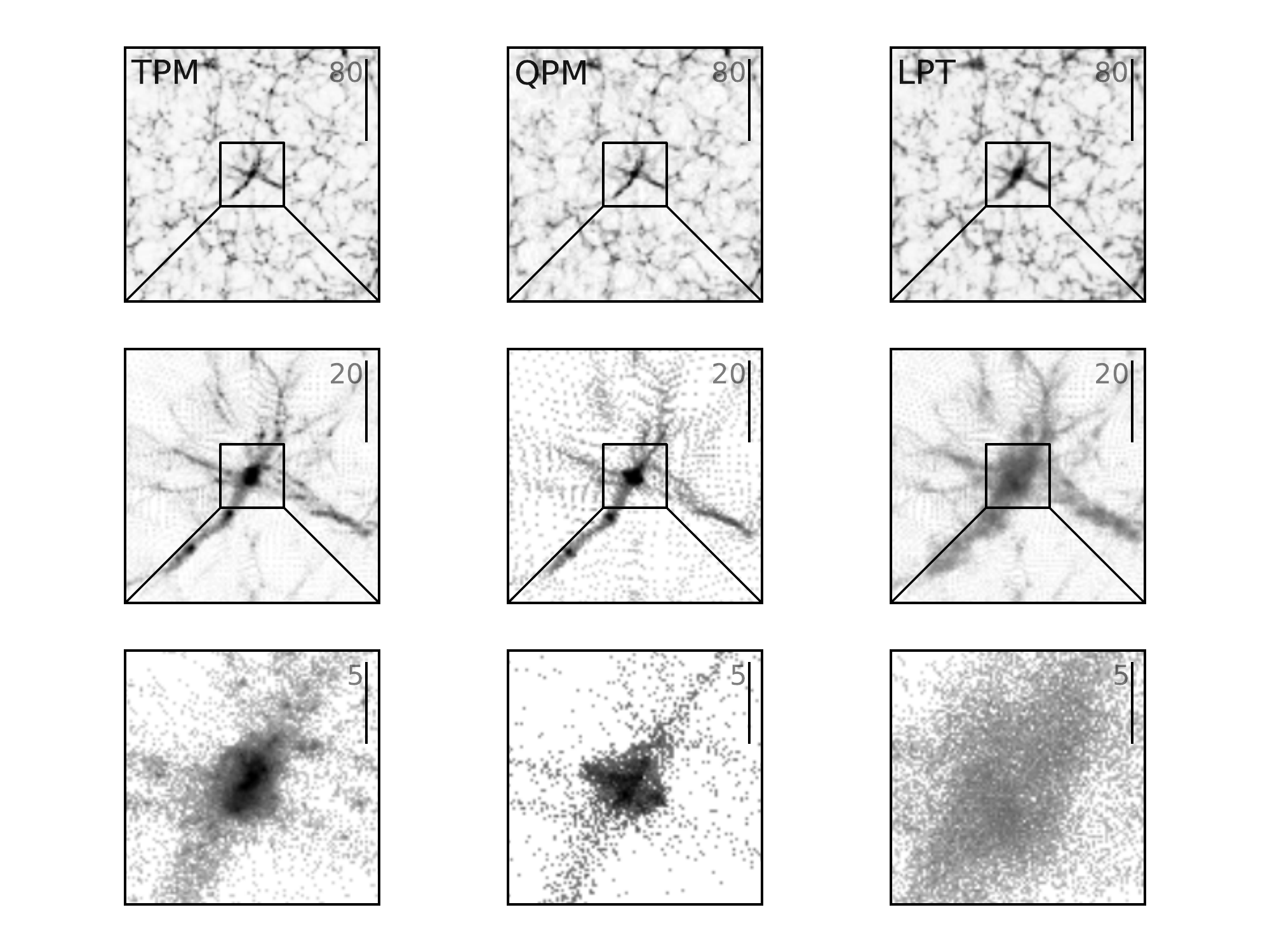}}
\end{center}
\caption{The density field in 3 different simulations with matched
initial conditions.  The grey scale is an arcsinh mapping of the density
(linear at low density and logarithmic at high density) in a slice
$4\,h^{-1}$Mpc think through a $2750\,h^{-1}$Mpc box.  The top row shows
a $256\times 256\,h^{-1}$Mpc region around the third most massive cluster
in the simulation ($M\simeq 1.6\times 10^{15}\,h^{-1}M_\odot$), the middle
row a zoom in to a $64\times 64\,h^{-1}$Mpc region and the bottom row a
further zoom in to a $16\times 16\,h^{-1}$Mpc region.
The three columns are for the TreePM simulation (TPM), the QPM simulation
(with $1/8$ the particle number) and the LPT simulation respectively.}
\label{fig:slice_mass}
\end{figure*}

In this section we compare several different methods of making mock catalogs
with those based upon the halos found in high resolution N-body simulations.
We shall treat the high resolution simulations as ``truth'' for the purposes
of the comparisons.


\subsection{Mass comparison}

While we will primarily be concerned with halo and galaxy catalogs in this
paper, we begin by comparing the matter fields produced by LPT and QPM with
that of the TPM simulation.  For simplicity we use one of the larger box
simulations and in making this comparison we ensure that all of the methods
use the same phases in the linear theory realization, so we expect the
structure to match in position across simulations.

A visual comparison of the density field produced by these three methods
is given in Fig.~\ref{fig:slice_mass}, which shows as a greyscale image the
mass density in a thin ($4\,h^{-1}$Mpc) slice through the box.  The slice
is centered on the $3^{\rm rd}$ most massive halo in the $2.75\,h^{-1}$Gpc
box and zooms in from a region $256\,h^{-1}$Mpc on a side to $16\,h^{-1}$Mpc
on a side.  The $256\times 256\times 4\,h^{-1}$Mpc slice (top row) shows that
all three methods produce the same filamentary structure on large scales (with
the random phases matched).  The $8\times$ coarser mass resolution in the
QPM (middle column) simulation compared to the TPM simulation (left column)
and the lack of small-scale power in the LPT simulation (right column)
is already evident in the middle row.  These trends are more apparent in the
bottom row, where we see the QPM code has merged a number of halos together
while the LPT code does not produce a bound structure at all.
In the original PThalos algorithm \citep{PThalos} the mass field was
``corrected'' by superposing the profiles of bound dark matter halos as
measured in high-resolution simulations.  In the LPT method of \citet{Man13}
this is not done (for the mass field) because the focus is on populating
halos with galaxies.  Were a more accurate mass field desired, it would be
straightforward to modify the algorithm to include this step.

\begin{figure}
\begin{center}
\resizebox{3.25in}{!}{\includegraphics{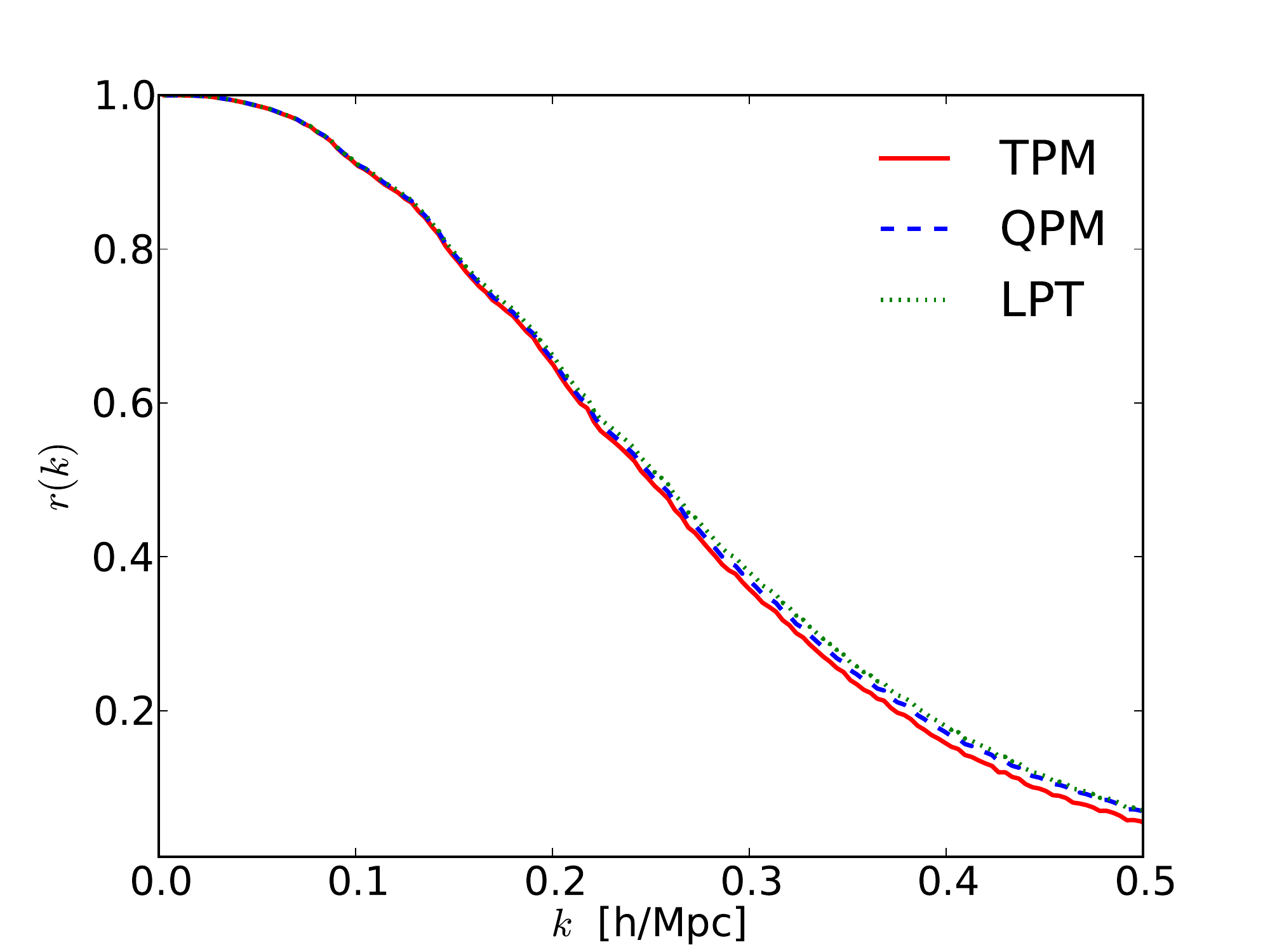}}
\end{center}
\caption{The Fourier-space cross-correlation coefficient,
$r\equiv P_{ij}/\sqrt{P_{ii}P_{jj}}$, between the initial conditions and
density field at $z\simeq 0.55$.
The lines are for TPM (red solid), QPM (blue dashed) and LPT (green dotted).
Note that the cross-correlation, sometimes called the ``propagator'', seen in
the TPM simulation is well reproduced by the QPM and LPT approximations on
all scales that are of interest here.}
\label{fig:mass_rk}
\end{figure}

In general the matter density field produced by the QPM simulation is highly
correlated with that of the TPM simulation on scales above the mesh scale.
The power spectra of the matter fields in the TPM and QPM simulations agree
to better than 5 per cent to $k\simeq 0.35\,h\,{\rm Mpc}^{-1}$ beyond which
the QPM simulation has less power than the TPM simulation\footnote{The
power spectrum of the LPT mass field has much less small-scale power, as
expected, and departs significantly from the TPM power spectrum by
$k\simeq 0.1\,h\,{\rm Mpc}^{-1}$.}.
The cross-correlation is above 95 per cent in Fourier space for
$k<1\,h\,{\rm Mpc}^{-1}$ and above 95 per cent in configuration space for
cubic cells larger than $2.7\,h^{-1}$Mpc (it falls to 87 per cent on the
mesh scale of the QPM simulation).
The cross-correlation\footnote{This is sometimes referred to as the
``propagator''.} between the initial density field and the field at
$z\simeq 0.55$ in the TPM simulations is also well reproduced by the QPM
and LPT runs, as shown in Fig.~\ref{fig:mass_rk}.

The distribution of the counts-in-cells is very similar for the QPM and
TPM runs on scales above $5\,h^{-1}$Mpc.  For the LPT runs there
is a pronounced lack of high density cells, as might be expected from
Fig.~\ref{fig:slice_mass}.  The discrepancy is an order of magnitude at
$\rho=10\,\bar{\rho}$ even for $10\,h^{-1}$Mpc cells.

\begin{figure}
\begin{center}
\resizebox{3.25in}{!}{\includegraphics{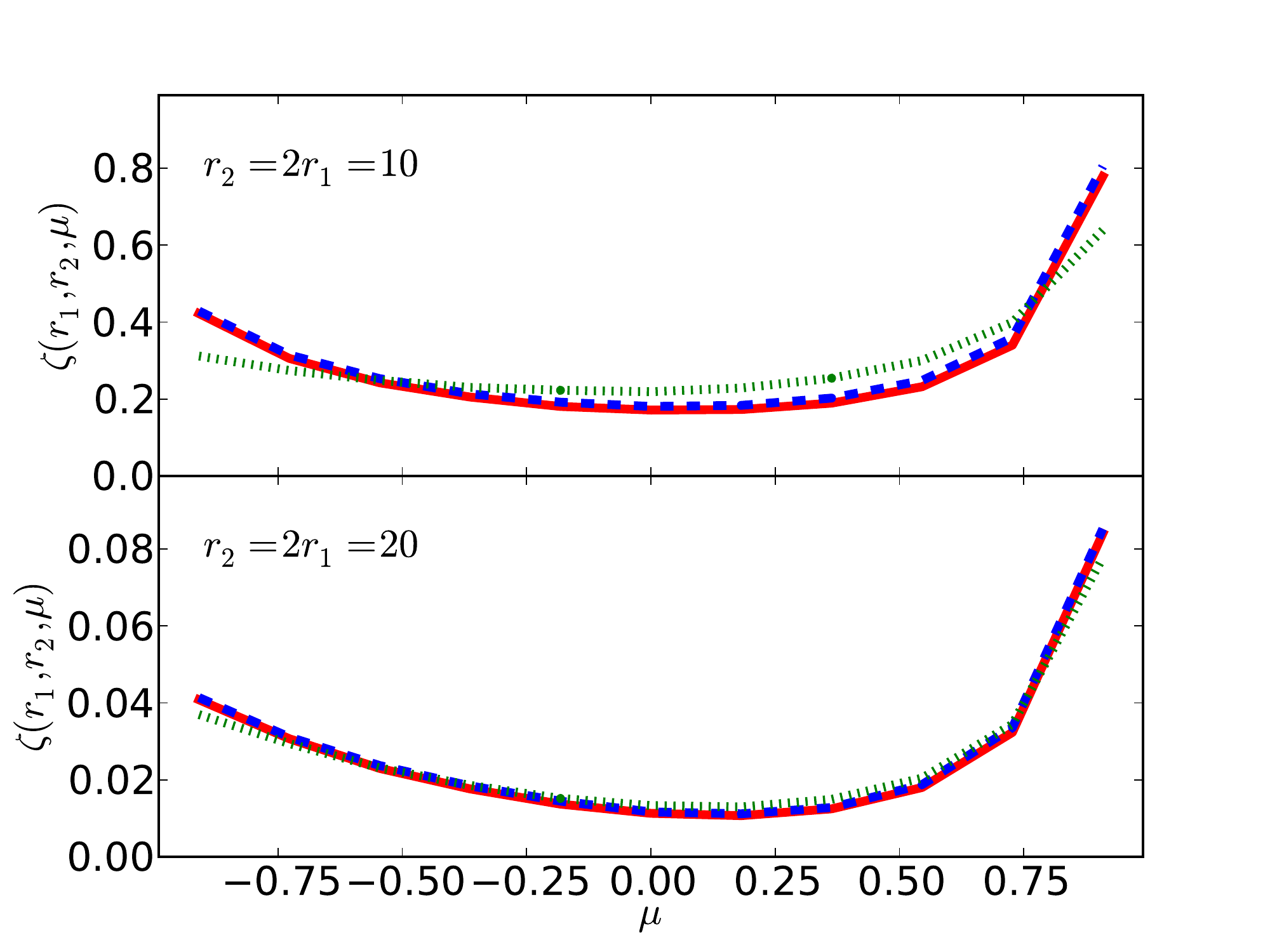}}
\end{center}
\caption{The real-space 3-point function of the matter density at
$z\simeq 0.55$ from the three simulations.  We show 3-point configurations
with side lengths in the ratio $r_2=2\,r_1$ as a function of the cosine of
the included angle ($\mu$).
The upper panel shows $r_2=10\,h^{-1}$Mpc and the lower panel
$r_2=20\,h^{-1}$Mpc.
The lines are for TPM (red solid), QPM (blue dashed) and LPT (green dotted).
Note that the matter distribution is more circular, or less filamentary,
on small scales for LPT than for QPM and TPM, but on large scales all
methods agree very well.}
\label{fig:mass3pt}
\end{figure}

One way of characterizing the filamentary nature of large-scale structure
is through the configuration dependence of the 3-point function (in real space).
As an example, the 3-point function for triangles with $r_2=2\,r_1$ is shown in
Fig.~\ref{fig:mass3pt}.  We see that the 3-point function of the QPM simulation
agrees very well with that of the TPM simulation on all scales plotted.
The LPT density field is also in good agreement with TPM on larger scales,
but appears to be slightly more circular (i.e.~less filamentary) for the
smaller triangles.

These comparisons suggest that the QPM simulation is well resolving the
large-scale structure in which the mock halos are to be placed, validating
its use as the input to the mock halo creation step.

\subsection{Halo comparison}

Each of our approximations provides us with a catalog of halo masses,
positions and velocities and galaxy (luminosities,) positions and
velocities.  While our primary interest is in the galaxy catalogs, we
first discuss the halos upon which they are built since such galaxy
catalogs are limited by the accuracy of the input halo catalogs.

\begin{figure}
\begin{center}
\resizebox{3.25in}{!}{\includegraphics{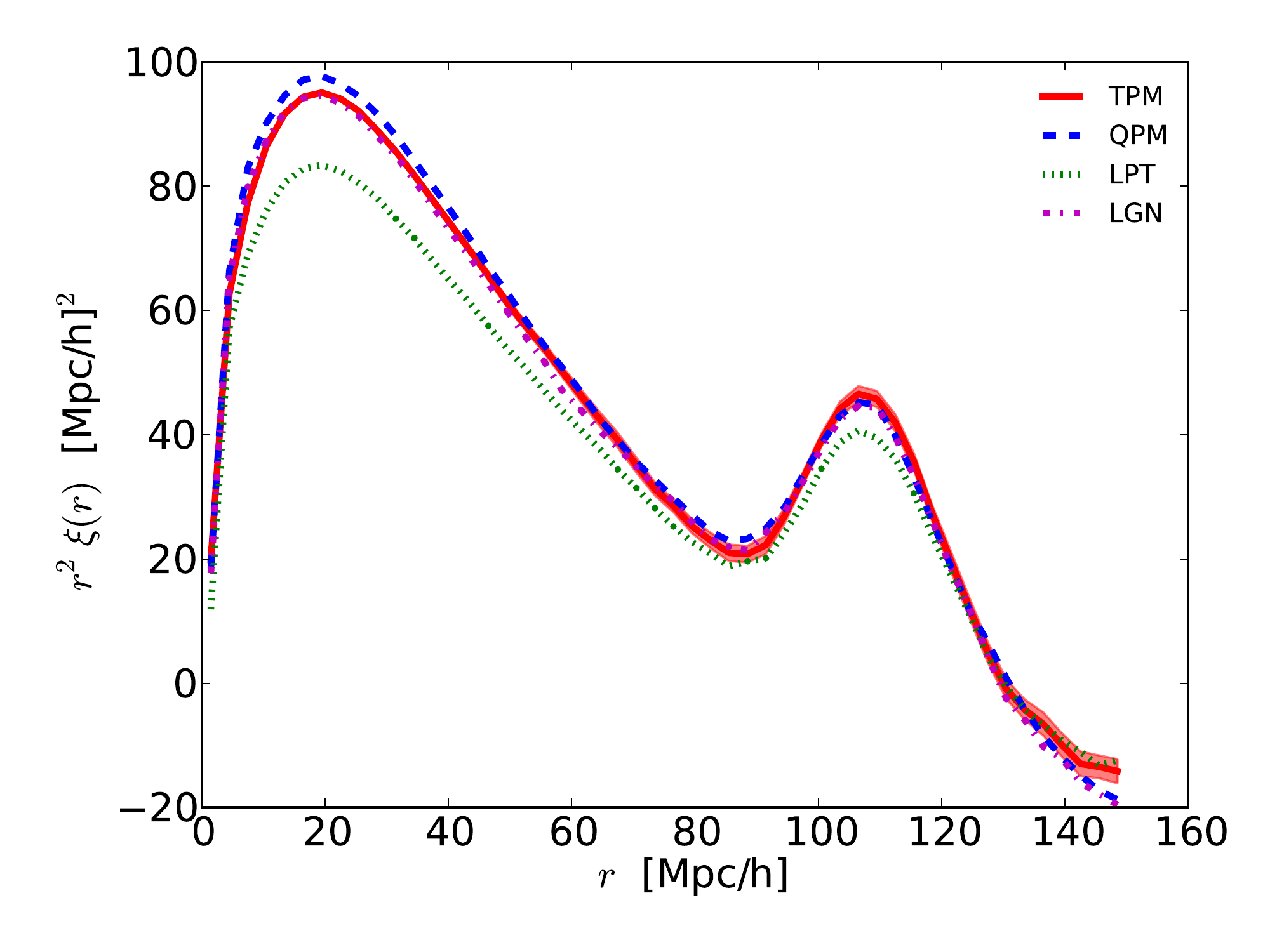}}
\end{center}
\caption{The real-space 2-point correlation function, $\xi(r)$,
for halos with $M>10^{13}\,h^{-1}M_\odot$.  The shaded red band shows the
$1\sigma$ error on the mean of the correlation function from the 20 TPM
mocks.  The uncertainty on the other lines
(QPM: dashed blue, LPT: dotted green, LGN: dot-dashed magenta)
is of the same size and is discussed further in the text.  The LGN
model almost perfectly overlaps the TPM $\xi(r)$, by construction.}
\label{fig:haloxir}
\end{figure}

The halo catalogs are necessarily sparser than the particles which defined
the mass density field, so in order to have better statistics across a wide
range of scales we make use of the 20 lower resolution simulations, with a
corresponding 20 mock simulations for each.  Throughout we shall plot the
average of each statistic over the 20 realizations to reduce sample variance.
Fig.~\ref{fig:haloxir} shows the real-space halo correlation function
(i.e.~configuration space 2-point function) for the 4 mock catalogs for a
sample of halos with $M>10^{13}\,h^{-1}M_\odot$, chosen to roughly match the
number density of BOSS CMASS galaxies \citep{Whi11,Aardvark,Daw13} though other
mass ranges look qualitatively similar.

\begin{figure}
\begin{center}
\resizebox{3.25in}{!}{\includegraphics{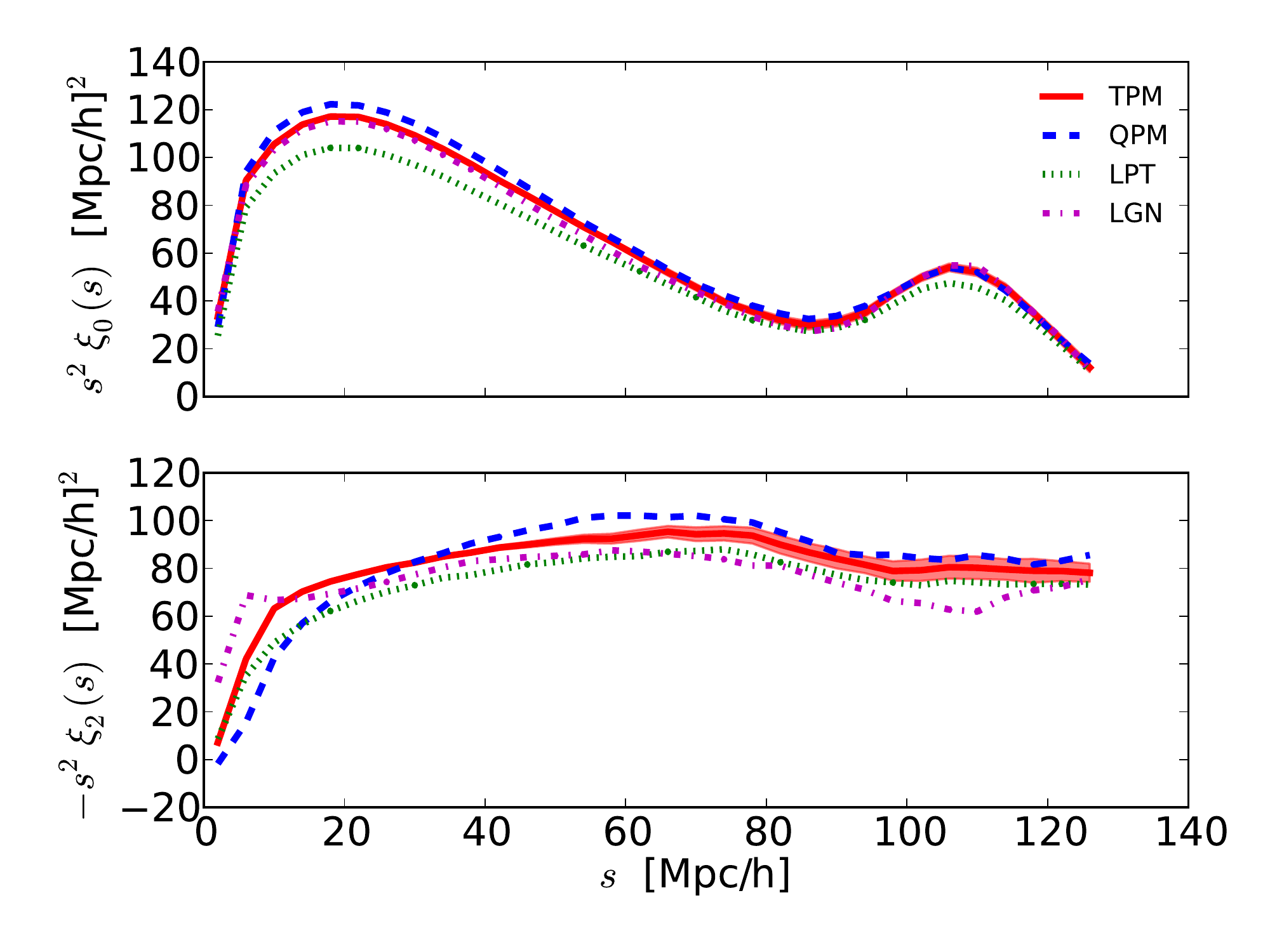}}
\end{center}
\caption{The reshift-space monopole (upper) and quadrupole (lower) 2-point
correlation functions for halos with $M>10^{13}\,h^{-1}M_\odot$.
The shaded red band shows the $1\sigma$ error on the mean of the correlation
function from the 20 TPM mocks.  The uncertainty on the other lines
(QPM: dashed blue, LPT: dotted green, LGN: dot-dashed magenta)
is of the same size and is discussed in the text.}
\label{fig:haloxis}
\end{figure}

\begin{figure}
\begin{center}
\resizebox{3.25in}{!}{\includegraphics{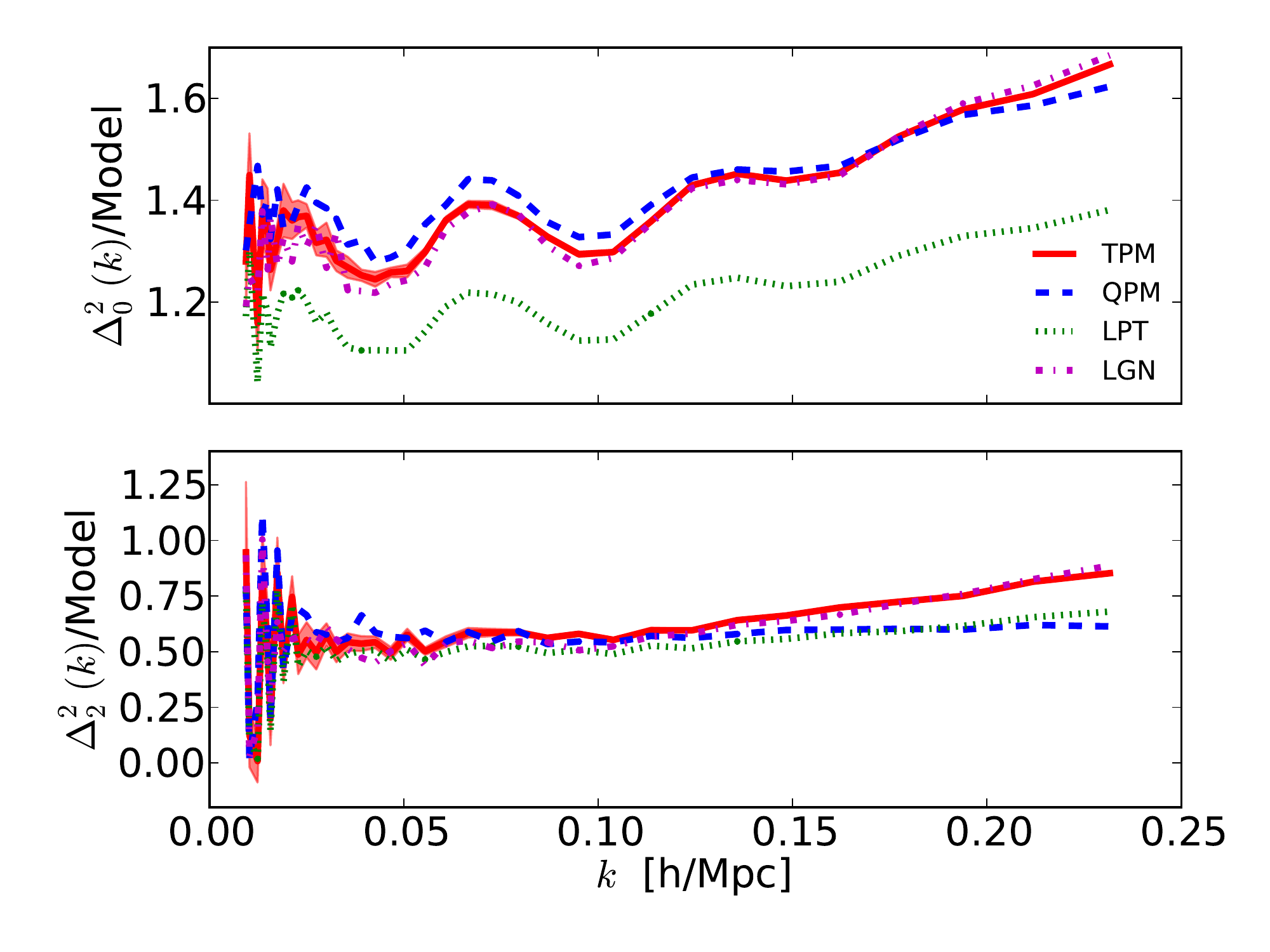}}
\end{center}
\caption{The reshift-space monopole (upper) and quadrupole (lower)
dimensionless power spectra for halos with $M>10^{13}\,h^{-1}M_\odot$.
We have divided each spectrum by a smooth model which is the linear theory,
real-space spectrum using the smooth fit to the linear theory transfer
function of \citet{EisHu98} and a scale-independent bias of $b=2$.
The amplitude thus reflects the effects of redshift-space distortions as well
as non-linearity, and the acoustic oscillations are highlighted.
The shaded red band shows the $1\sigma$ error on the mean of the spectra from
the 20 TPM mocks.  The uncertainty on the other lines
(QPM: dashed blue, LPT: dotted green, LGN: dot-dashed magenta)
is of the same size.}
\label{fig:halopks}
\end{figure}

\begin{figure}
\begin{center}
\resizebox{3.25in}{!}{\includegraphics{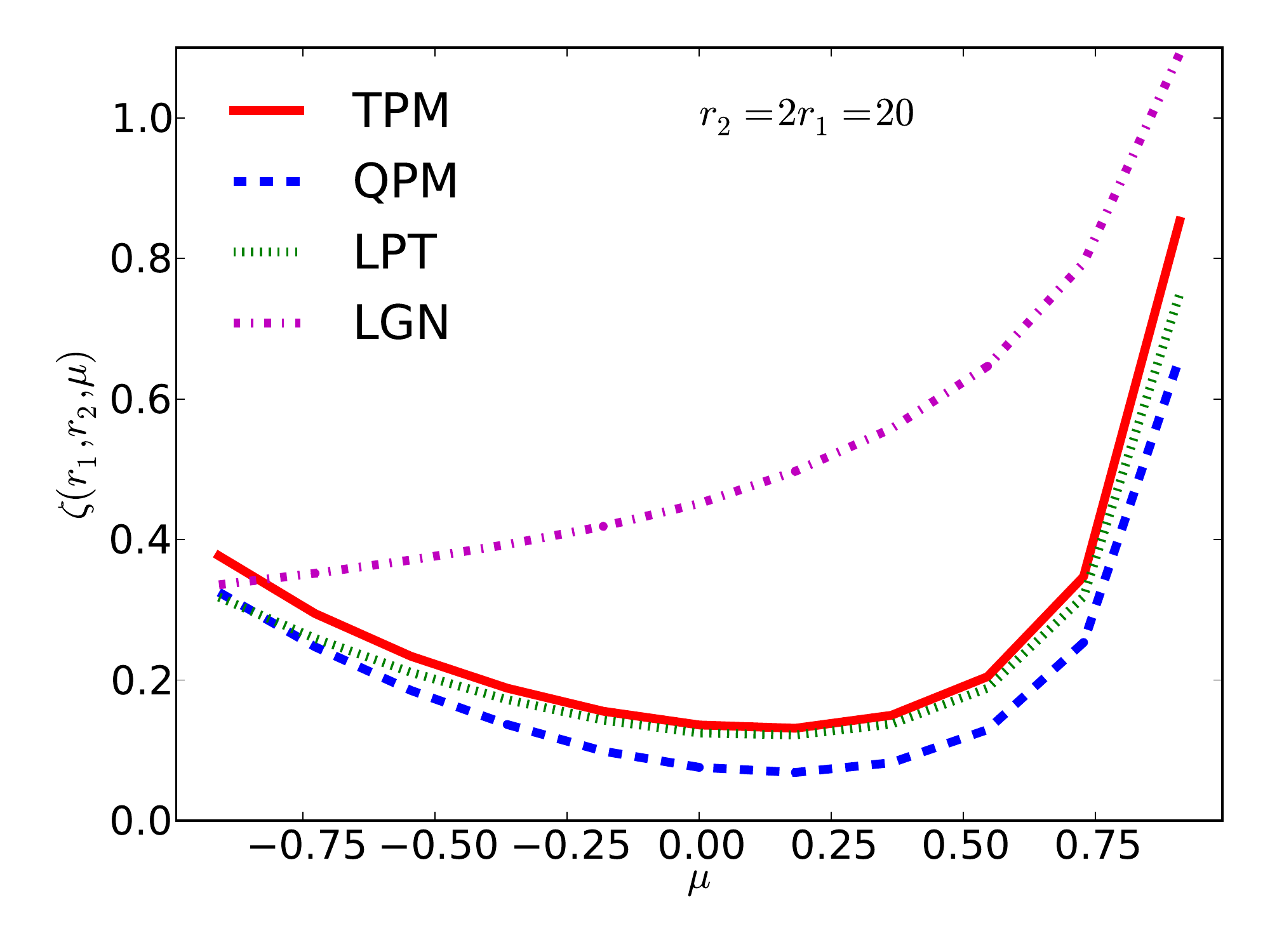}}
\end{center}
\caption{The real-space 3-point function of the halo density at $z\simeq 0.55$
{}from the four mock catalogs.  We show 3-point configurations with side
lengths $r_2=2\,r_1=20\,h^{-1}$Mpc as a function of the cosine of the included
angle ($\mu$).  The agreement between the TPM (solid red) and
LPT (dotted green) mocks is very good on these scales, and slightly better
than the agreement between TPM and QPM (dashed blue).  The LGN
mocks (dot-dashed magenta) are clear outliers.}
\label{fig:halo3pt}
\end{figure}

Fig.~\ref{fig:haloxir} shows that the large-scale bias of the halos produced
by all of the methods is approximately correct.  (We are not concerned here
with the level of agreement below $10\,h^{-1}$Mpc scales, though we shall come
back to that in the next section.)  On large scales we see that the LGN
mocks agree almost perfectly with the TPM mocks, as expected.  Those mocks
are designed to produce a given 2-point function, in this case produced using
convolution Lagrangian perturbation theory \citep[CLPT;][]{CLPT}.  Since CLPT
provides a good fit to the halo correlation function we fully expected our
LGN mocks to fit as well, and this expectation is born out.
The QPM mocks provide an adequate fit to the real-space correlation function,
with a slight excess clustering near $20\,h^{-1}$Mpc.  This excess appears
visually quite significant due to the choice of $r^2\xi$ as the y-axis, but
is only 3 per cent in $\xi$ at $20\,h^{-1}$Mpc and is completely acceptable
for our purposes. 
Finally we see that the LPT mocks tend to undershoot the correlations over a
wide range of scales and have a slightly different shape near the acoustic
peak ($r\sim 100\,h^{-1}$Mpc).  This is due to our choice of a broad range
of halos masses and a single linking length.  For a narrower range of halos
we can adjust the linking length to obtain a better match, but we found it
difficult to simultaneously match the bias and mass function across a wide
range of halo masses with a single value of the linking length.

The situation in redshift-space is shown in Fig.~\ref{fig:haloxis} which
presents the monopole and quadrupole moments of the correlation function.
For the monopole all of the mocks agree relatively well.
Once again the LPT mocks show a slight deficit of clustering across a
broad range of scales, the QPM mocks overshoot near $20\,h^{-1}$Mpc and
the LGN mocks match very well.
For the quadrupole the agreement is less good, but still quite acceptable
given that the observations tend to have larger errors on the quadrupole
than the monopole.

An alternate view of the redshift-space statistics is given in
Fig.~\ref{fig:halopks}, which presents the monopole and quadrupole
power spectra divided by a smooth model. The trends are very similar
to those in Fig.~\ref{fig:haloxis}.

Figure \ref{fig:halo3pt} shows the configuration-space 3-point correlation
function for triangles with $r_2=2\,r_1=20\,h^{-1}$Mpc as a function of the
cosine of the included angle.  The TPM and LPT mocks agree quite well in
the amplitude and shape of the 3-point function.  The QPM mocks agree slightly
less well, but still have approximately the right shape, indicating the
prevalence of filaments in the cosmic web.
(Note that $10-20\,h^{-1}$Mpc are the scales where the QPM mocks agree least
well with the TPM simulation for the two point function.)
The clear outlier is the LGN mocks, where the remapping of the original
Gaussian density field does not produce the same filamentary, beaded web as is
produced by gravitational evolution and the shape dependence of the 3-point
function is very different from that of the other mocks.
{}From this we infer that laying down halos based on a Gaussian field does not
produce the right structure in the halo density field.
Laying down halos based on the non-linearly evolved field does a slightly
better job, but since the mock halos are selected at random based only on
local density the filamentary nature of the halo distribution is slightly
washed out.
Finding overdense regions in the evolved field does best of all, but is the
most expensive in terms of memory or computational cost.

\begin{figure}
\begin{center}
\resizebox{3.25in}{!}{\includegraphics{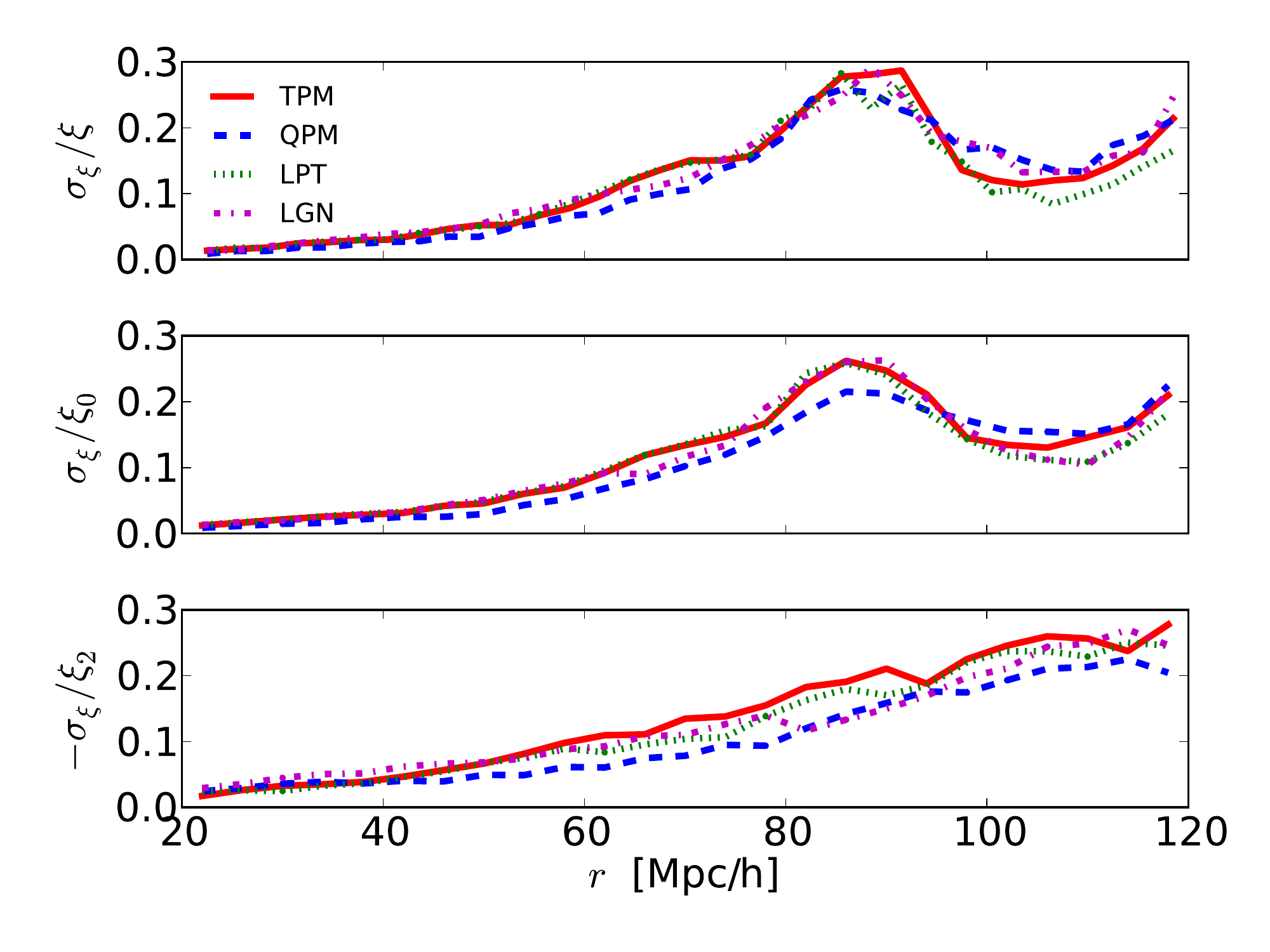}}
\end{center}
\caption{The fractional uncertainty in the real-space halo correlation
  function (upper) or monopole (middle) or quadrupole (lower) of the
  redshift-space correlation function as estimated from the 20 mock
  catalogs of each type.  {}From 20 realizations we expect the error
  on the error to be about 30 per cent, so all of these methods agree
  within errors.}
\label{fig:halo_xi_err}
\end{figure}

Figure \ref{fig:halo_xi_err} shows the fractional uncertainty on the
correlation function, computed from the 20 independent simulations, for
each of the methods.  We expect linearly biased tracers of the matter field
to behave in a similar fashion on large scales, and that is born out in this
comparison.
The trends for the real- and redshift-space statistics are very similar,
and largely agree between the 4 mock methods.
Recall that with only 20 simulations, if we approximate the individual
measurements as Gaussian we would expect an error on our fractional error
of 30 per cent, so we are unable to statistically distinguish between these
approximate methods.  Unfortunately it is too expensive to increase the
number of TPM simulations to reduce the error and provide a stronger test.
Near the acoustic peak the fractional error on the real-space correlation
function for this volume and number density, assuming linearly biased linear
theory and Gaussian statistics, is in good agreement with the numbers shown
in Fig.~\ref{fig:halo_xi_err} and the error is almost totally dominated by
sample variance.  Given that all of the methods are doing a good job of
modeling the largest scales which dominate the sample variance, it is not
unreasonable to expect they produce the same scatter in the correlation
function.

\subsection{Galaxy comparison} \label{sec:galaxycompare}

From the halo catalogs\footnote{Since the log-normal mocks don't
  produce a halo catalog with assigned masses, we don't include them
  in this section.  The way in which log-normal mocks are typically
  used is to directly produce the galaxies from the log-normal density
  field, rather than going through an intermediate step.  The
  comparison is then qualitatively similar to that for the halos which
  we discussed above and we gain little insight by repeating it here.}
we create galaxy catalogs using a halo-occupation-distribution
approach \citep{PeaSmi00,Sel00,Ben00,Sco01,WHS01,BerWei02,CooShe02}.  We use
the HOD parameterization of \cite{Tin12}.  Once a set of HOD parameter
values has been chosen, we populate each mock halo in a given output
with mock ``galaxies''.  The HOD provides the probabilities that a
halo will contain a central galaxy and the number of satellites.  The
central galaxy is placed at the center of the halo (which is the
position of the most bound particle in the halo for the FoF halos and
the position of the particle for the mock halos), and we place
satellites assuming a spherical \citet{NFW} profile with a
concentration determined using the method of \citet{Mun10}.

\begin{figure}
\begin{center}
\resizebox{3.25in}{!}{\includegraphics{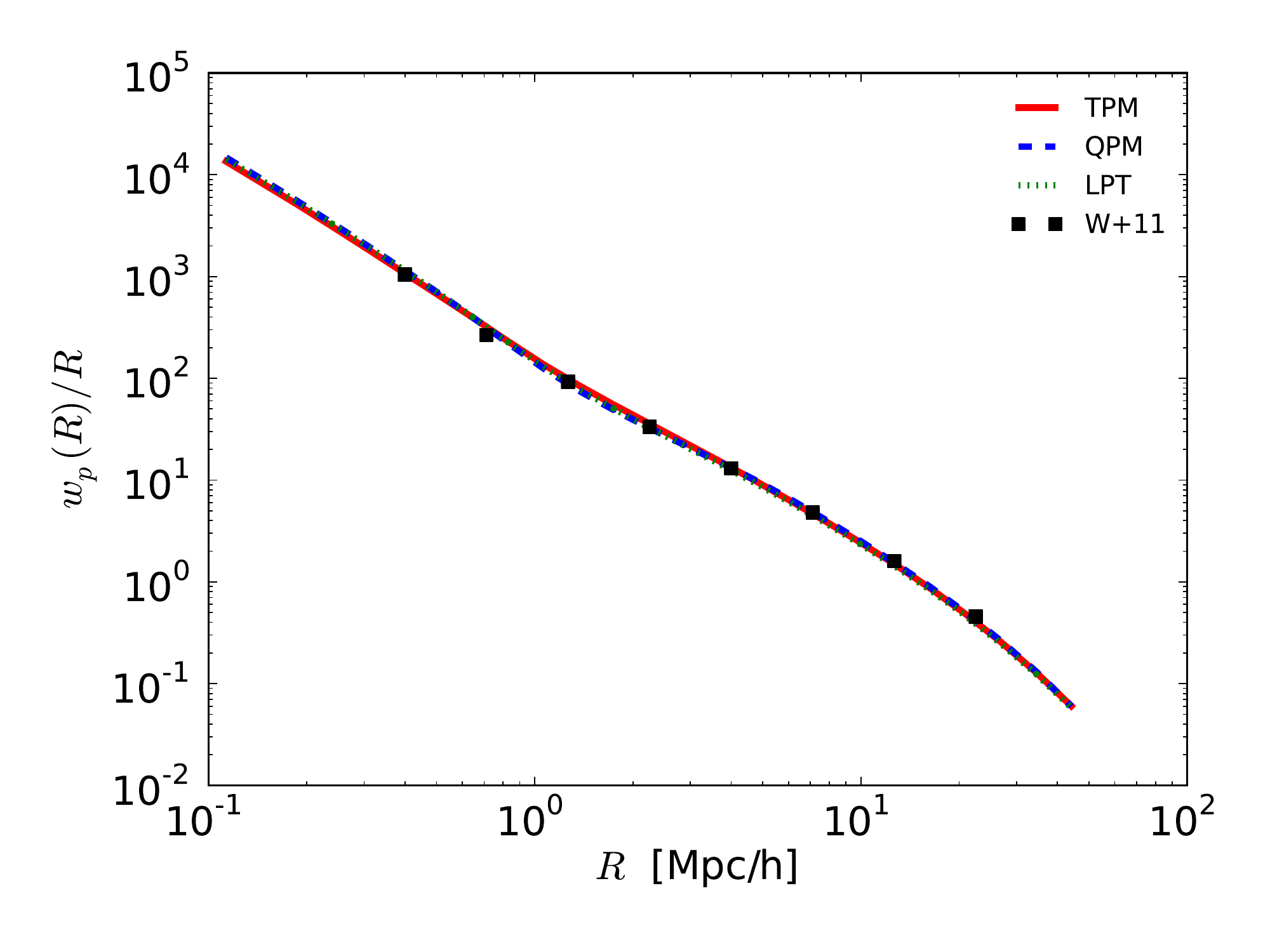}}
\end{center}
\caption{The projected correlation function, $w_p(R)$, as a function of
projected radius, $R$, for mock galaxies.  For each of the mock types, the
HOD has been independently adjusted to fit the data.  The lines indicate the
best-fit prediction from the halo catalog + HOD modeling while the points
are the measurements of \protect\citet{Whi11} (with the errors suppressed for
clarity).  In each case the models provide an excellent fit to the small-scale,
real-space clustering -- the lines lie almost on top of each other.}
\label{fig:galaxy_wp}
\end{figure}

We follow standard practice and adjust the HOD parameters for each mock to
fit the small-scale projected two-point clustering.  As an illustrative example 
we have chosen to use the projected correlation function of BOSS CMASS galaxies
reported in \citet{Whi11}.  In each case we are able to match the clustering
between the mocks and the observational measurements, as shown in 
Fig.~\ref{fig:galaxy_wp}.

\begin{figure}
\begin{center}
\resizebox{3.25in}{!}{\includegraphics{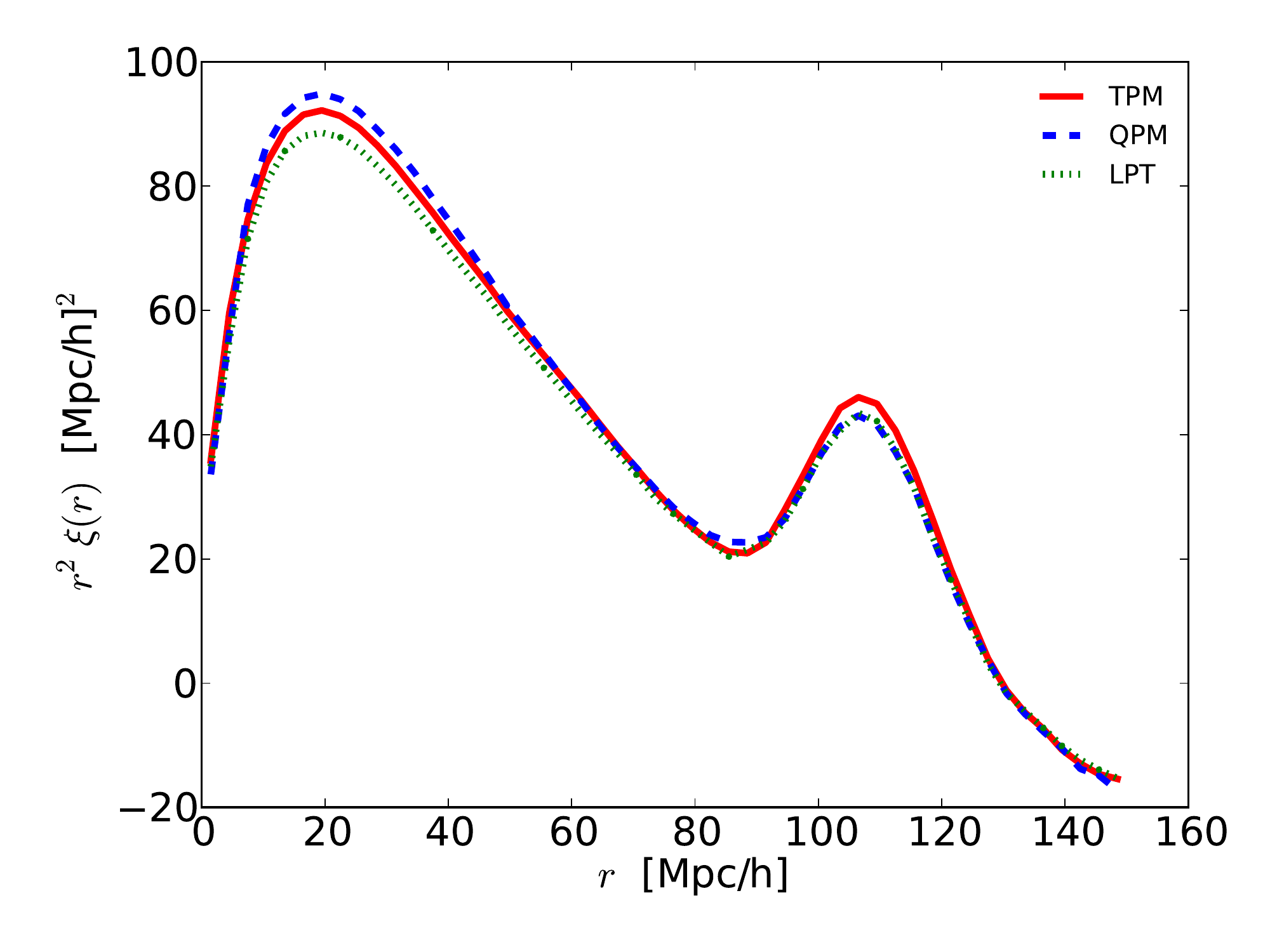}}
\end{center}
\caption{The real-space 2-point correlation function, $\xi(r)$,
for mock galaxies.}
\label{fig:galaxy_xir}
\end{figure}

The real-space correlation function for each of the mock galaxy catalogs is
shown in Fig.~\ref{fig:galaxy_xir} which is the analog of
Fig.~\ref{fig:haloxir}.  As in that case we see that the mocks agree with the
high-resolution N-body simulation very well over a broad range of scales.
The QPM mocks have slightly more power near $20\,h^{-1}$Mpc but only at the
few per cent level.
This indicates that we are placing halos correctly, in a statistical sense,
within the large-scale mass field and any higher order correlations between
halos that we are missing by such a random assignment don't impact the
low-order galaxy statistics.

\begin{figure}
\begin{center}
\resizebox{3.25in}{!}{\includegraphics{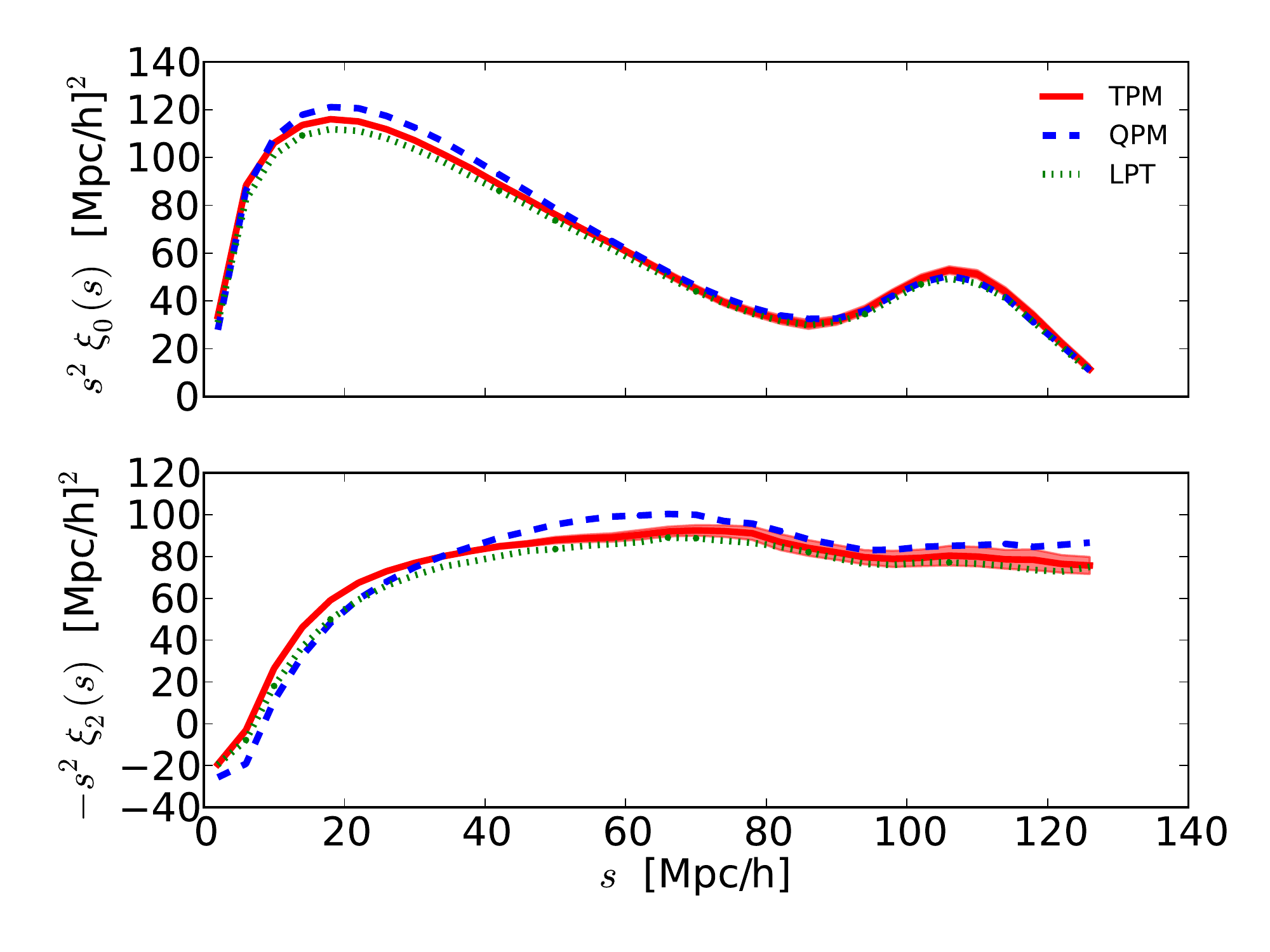}}
\end{center}
\caption{The reshift-space monopole (upper) and quadrupole (lower) 2-point
correlation functions for mock galaxies.
The shaded red band shows the $1\sigma$ error on the mean of the correlation
function from the 20 TPM mocks.  The uncertainty on the other lines
(QPM: dashed blue, LPT: dotted green)
is of the same size and is discussed in the text.}
\label{fig:galaxy_xis}
\end{figure}

\begin{figure}
\begin{center}
\resizebox{3.25in}{!}{\includegraphics{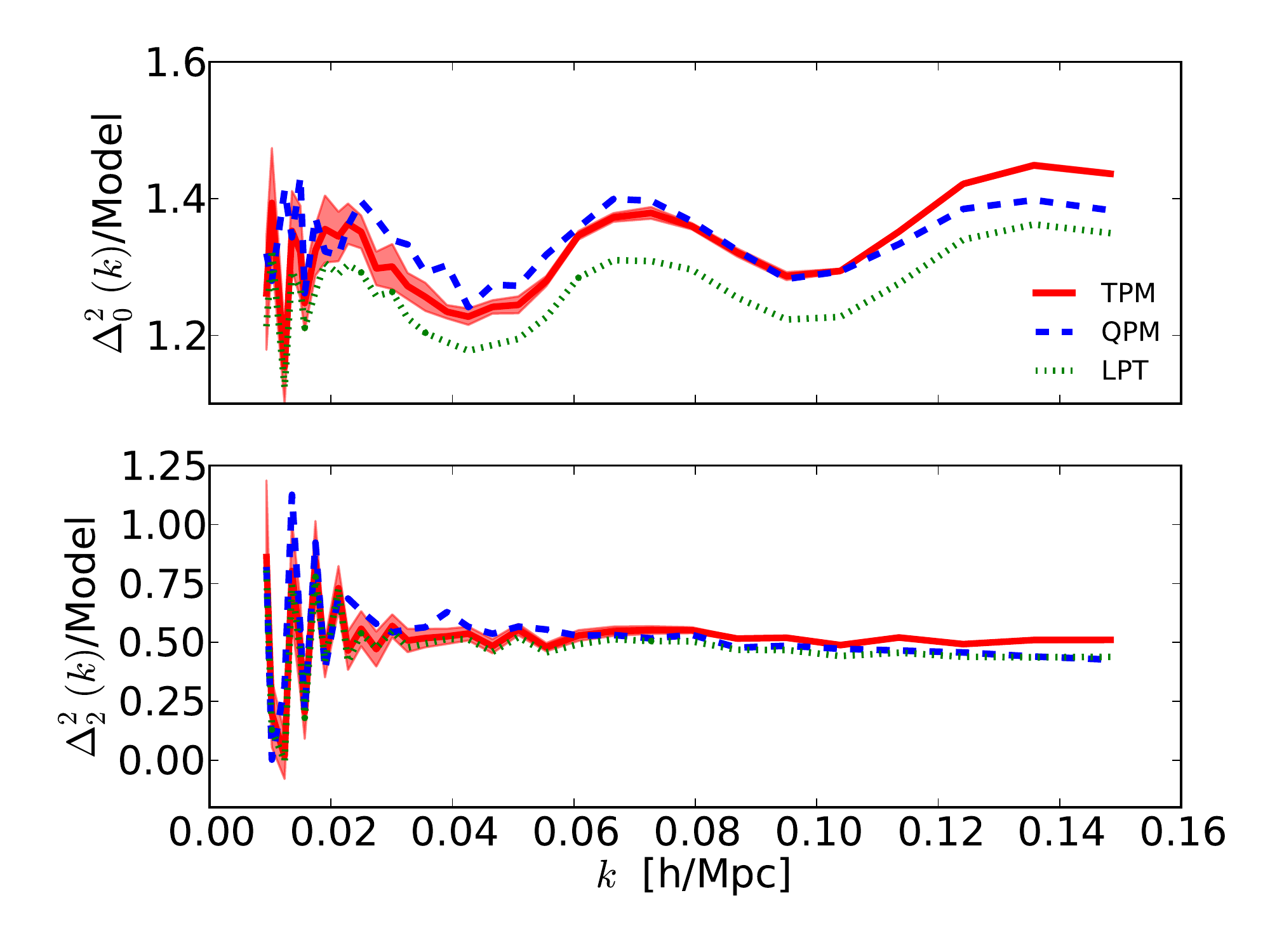}}
\end{center}
\caption{The reshift-space monopole (upper) and quadrupole (lower)
dimensionless power spectra for mock galaxies.
We have divided each spectrum by a smooth model which is the linear theory,
real-space spectrum using the smooth fit to the linear theory transfer
function of \citet{EisHu98} and a scale-independent bias of $b=2$.
The amplitude thus reflects the effects of redshift-space distortions as well
as non-linearity, and the acoustic oscillations are highlighted.
The shaded red band shows the $1\sigma$ error on the mean of the spectra from
the 20 TPM mocks.  The uncertainty on the other lines
(QPM: dashed blue, LPT: dotted green)
is of the same size.}
\label{fig:galaxy_pks}
\end{figure}

The situation in redshift-space is shown in Fig.~\ref{fig:galaxy_xis} which
is the analog of Fig.~\ref{fig:haloxis} and presents the monopole and
quadrupole moments of the correlation function.  The level of agreement is
very much as we saw in Fig.~\ref{fig:haloxis}, indicating that we are
modeling fingers-of-god and satellite fractions comparably in all three cases.
The complementary view is given in Fig.~\ref{fig:galaxy_pks}, which shows the
monopole and quadrupole of the redshift-space power spectrum.

\begin{figure}
\begin{center}
\resizebox{3.25in}{!}{\includegraphics{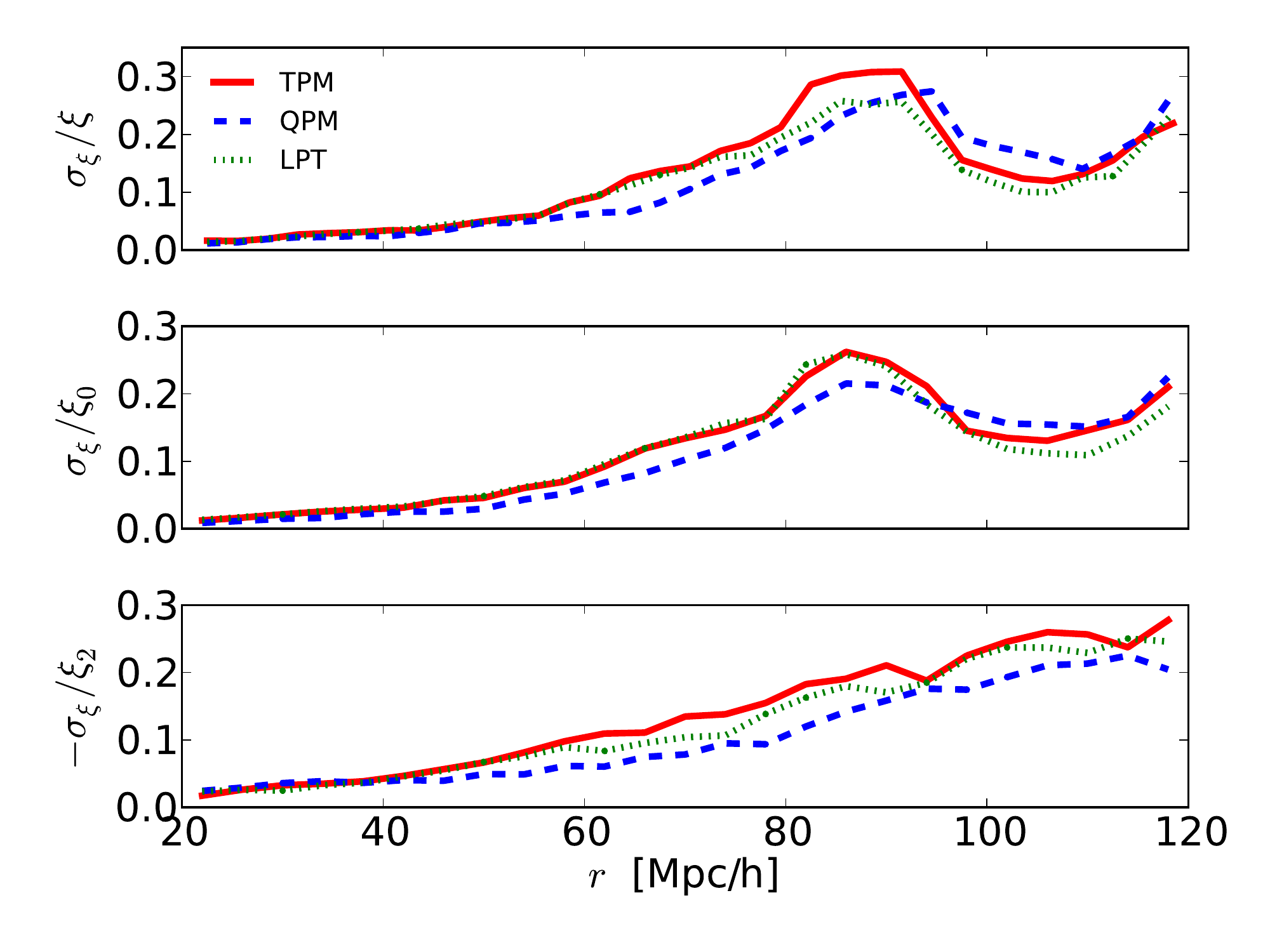}}
\end{center}
\caption{The fractional uncertainty in the real-space correlation function
(upper) or monopole (middle) or quadrupole (lower) of the redshift-space
correlation function as estimated from the 20 mock catalogs of each type.
{}From 20 realizations we expect the error on the error to be about 30
per cent, so we are unable to statistically distinguish these different
approximations from the simulations we have.}
\label{fig:galaxy_xi_err}
\end{figure}

Finally we show the fractional uncertainty in the real-space, monopole
and quadrupole correlation functions vs.~scale in Fig.~\ref{fig:galaxy_xi_err}.
As in Fig.~\ref{fig:halo_xi_err} the small number of high-resolution N-body
simulations means we only have a $\sim 30$ per cent prediction of the
fractional error with which to compare.  Within this uncertainty the methods
agree well, as expected since they are approximately linear tracers of the
large scale matter field.

\subsection{Computational cost}

For each of the mock catalog methods described above the most time consuming 
steps involve Fast Fourier Transforms.  We use the parallel, real Fourier 
transforms in the FFTW package\footnote{www.fftw.org}.
The FFT scales with problem size as $N\log N$, so that doubling the
linear dimension of the grid increases the run time by approximately
an order of magnitude (in addition to requiring about an order of magnitude
more memory).  The additional cost of moving the particles is small,
though with more particles and more clustered particles the time taken to
find halos with the FoF algorithm and halo properties (for those cases
where this is a step in the process) increases and can become a
significant fraction of the total runtime for some methods.

As each of the methods has the same underlying operation as its most
computationally demanding step, we can roughly compare ``cost'' by
simply comparing the number of FFTs used for each.  The LGN
mocks require 4 FFTs (one to convert from $\delta(k)$ to $\delta(x)$
and then 3 to compute the displacements).  In our implementation the
LPT mocks require 13 FFTs.  With our fixed time steps the QPM code
takes about 20-30 time steps and requires 2 FFTs per step (plus 13
FFTs for the generation of the initial conditions).  Since we use a
lower resolution with fewer particles for the QPM run however each
step is about $8\times$ faster and requires less memory than the
LGN or LPT mocks.  For the same particle number then the
LGN simulations run a few times faster than the LPT
simulations, which are comparable to the QPM simulations which is more
than two orders of magnitude faster than the TPM simulation.

\section{Mock Factory} \label{sec:mockFactory}

So far we have dealt with producing a periodic cube populated with mock
galaxies according to a specified HOD.  In many instances it is also
important to reproduce other aspects of the data, such as applying 
redshift-space distortions and trimming to the survey geometry. 
Our implementation of these steps has three separate parts, any of which 
can be used independently: 
(1) creation of mock halo files from local densities, 
(2) application of an HOD to map mock galaxies to halos, and
(3) transformation of the periodic box of objects into a realistic catalog
on the sky.
We make this code, referred to as \texttt{mockFactory}, publicly 
available\footnote{\texttt{http://github.com/mockFactory}}. 

The first step was the topic of the previous sections.  The code
selects, from a random sample of dark matter particles with densities,
a set of mock halos with masses.  The halos match a specified mass
function and large-scale bias vs.~mass relation, for which we use the
fits of \citet{Tin08,Tin10} for overdensity $\Delta_\rho=200$ halos.  The
result of this stage is a periodic box of halos, stored as a
\texttt{HaloFile}.  The operation of the code is controlled through a
parameter file, which is ascii text. The performance and memory
considerations of the code depend on the size of the PM simulation and
the fraction of particles that are retained in the subsample file from
which the mock halos are culled. For test runs on a PM simulation with
$1280^3$ particles, of which 10\% were retained in the subsample file,
the run-time on a typical workstation core is $\sim 50$ minutes to
create the mock halos, requiring $5.5\,$GB in memory (effectively no
overhead relative to the size of the input subsample file). The run
time scales linearly with the size of the subsample file.

The second part of the code populates a halo catalog (typically the
\texttt{HaloFile} described above, but it is possible to use any halo
file provided it is in the proper format) with artificial galaxies.
This applies a HOD as described in Section \ref{sec:galaxycompare}.  The 
central galaxy is placed at the center of the halo and satellites are 
placed assuming a spherical \citet{NFW} profile with a concentration 
determined using the method of \citet{Mun10}, which can be scaled by a 
factor in the input parameter file.  In the examples used in this paper, 
we use the central-satellite HOD parameterization used in \cite{Tin12}, 
although the code has a number of options for these parameterizations. 
The code can also adjust the input HOD parameters to match a desired 
galaxy number density by rescaling all the halo mass values by the same 
factor. To fill the \texttt{HaloFile} with mock galaxies requires 
substantially less time and memory: approximately 3 minutes for the 
example introduced above and only the \texttt{HaloFile} is held is 
memory.

The third part of the code, called \texttt{make\_survey}, takes a periodic
box of objects (mock galaxies, quasars, halos or particles from a simulation
box) and projects them on the sky, optionally applying various layers of 
realism to represent a mock survey.
Our implementation can perform a number of steps which include: 
\begin{itemize}
\item[(a)] volume remapping of the periodic box using BoxRemap \citep{CarWhi10}
     \footnote{\texttt{http://mwhite.berkeley.edu/BoxRemap/}}
\item[(b)] box translation and rotation, 
\item[(c)] sky projection using cosmological distances, 
\item[(d)] modeling redshift distortions using peculiar velocities, 
\item[(e)] trimming to a survey footprint using a MANGLE mask \citep{Swa08},
\item[(f)] downsampling based on angular sky completeness, and 
\item[(g)] downsampling based on radial selection.
\end{itemize}
A configuration file controls which of these steps are applied, as well as
giving any required parameters. In addition, several tools exist within 
the code base to assist the user in determining and testing the parameters.
The \texttt{make\_survey} code is written to be relatively efficient.
It reads the input catalog one object at a time, processes each object
through all required steps, and outputs any object that makes all cuts
before moving on to the next.  This results in minimal memory usage as
the input catalog is \emph{never} fully read into memory, nor is the
output stored.
The runtime is sufficiently fast such that it should not require significant
computing resources even for a large number of mock catalogs.

\section{Summary}
\label{sec:summary}

In the rising age of large-scale galaxy redshift surveys, the need for
fast and accurate methods to create mock galaxy catalogs is at a
premium. In this paper we have presented a new method for creating
such mocks, which we call the ``quick particle mesh'' method
(QPM). Our method is based on using rapid, low-resolution particle
mesh simulations that accurately reproduce the large-scale dark matter
density field. Particles are sampled from the density field based on
their local density such that the sampled particles have $N$-point
statistics nearly equivalent to the halos resolved in high-resolution
simulations. Thus our method creates a set of mock halos that can be
populated using halo occupation methods to create galaxy mocks for a
variety of possible target classes.

We compare the real- and redshift-space clustering statistics of our
mock halos and mock galaxies to those of second-order Lagrange
perturbation theory (LPT) and the log-normal method (LGN). These two
methods are currently the most widely used for creating large sets of
mocks. We use high-resolution N-body simulations (TPM) as ``truth'' in
these comparisons, which focus on halo mass ranges applicable to those
probed by BOSS-type samples. The halo clustering
produced by the QPM method agrees quite well with TPM results. For two-point
statistics, our implementation of LPT slightly underpredicts the halo
clustering. The LGN method offers an excellent match to the TPM halo
clustering, but cannot reproduce the behavior of the density field for
higher-order statistics.

There are several benefits of our method compared to other current
methods.  In comparison to some perturbation-theory methods, QPM
offers improvement in terms of both run-time and memory requirements.
Although a PM simulation performs more FFTs than LPT, with our particle
sampling scheme the grid size required to achieve the same halo mass
resolution is less, thus the transforms run significantly faster
and require less memory.
Although we restrict our comparisons to halo mass ranges $\gtrsim 10^{13}
h^{-1}M_\odot$, the lower mass limit of the QPM \texttt{HaloFile} can
be set to any required value. Increasing the mass resolution of LPT
requires increasing the size of the density grid on which the FFTs are
performed.  We note, however, that our current implementation of QPM is
designed for BOSS-type galaxy samples, thus achieving accurate halo
catalogs at $M_{\rm halo} \lesssim 10^{12} h^{-1}M_\odot$ requires
additional calibration.
When comparing TPM to the LGN catalogs we find the LGN method reproduces
the two-point clustering better than either QPM or LPT.
The filamentary nature of the field is not well reproduced however.
In addition, a single QPM \texttt{HaloFile} can be used to create galaxy
mocks for a variety of targets with varying bias values; the LGN simulation
is tuned to a specific value of bias, thus must be re-run for any variation of
target class.  In some situations this is a disadvantage.

The QPM method, as discussed here, uses a simple scheme of determining 
mock halos from local particle densities.
While successful, there were a few discrepancies in the halo clustering
that were not present in the mass fields.
It is possible that an improved mapping could be found, perhaps using more
information from the density field (or initial conditions) or conditionally
selecting particles based on previous ``draws'', and this would mitigate some
of these discrepancies.  We have not explored this avenue here as we do not
believe the discrepancies to be a serious limitation at present --
the differences are small and often reduced further once mapped to mock 
galaxies. 

We have released our QPM implementation as part of the \texttt{mockFactory} 
software package.  Users supply their own PM simulations, and the code can
create the mock halo file, create the mock galaxy catalog, and convert
this galaxy catalog from a periodic cube to a cut-sky angular mock in
redshift space.
We do not supply the PM code, but there are many public codes that have
been thoroughly tested and used throughout the community: e.g.~the code of
\cite{Kly97}\footnote{http://astro.nmsu.edu/$\sim$aklypin/pmcode.html},
PMFAST\footnote{http://www.cita.utoronto.ca/$\sim$merz/pmfast/}
(\citealt{Mer05}),
as well as tree-particle-mesh codes that can be run as PM-only, such as
GADGET2\footnote{http://www.mpa-garching.mpg.de/gadget/}
(\citealt{Spr05}) and the TPM
code\footnote{http://www.astro.princeton.edu/$\sim$bode/TPM/}
of \cite{Bod03}.
Additionally, if the reader is sufficiently motivated, instructions for
creating one's own PM code are also
available\footnote{http://astro.uchicago.edu/$\sim$andrey/talks/PM/pm\_slides.pdf}.

QPM offers a method that is fast, requires relative little memory, and is
highly flexible.  It's low computational requirements and high flexibility
make QPM optimal for the next generation of large-scale structure surveys
that will push the limits of current methods both in terms of the volume
they probe and the bias of the galaxies they target.

\section*{Acknowledgments}

The simulations used in this paper were analyzed at the
National Energy Research Scientific Computing Center.
MW is supported by the NSF and NASA.

\appendix

\section{Mock catalogs} \label{app:mocks}

We have compared the QPM method described in this paper to a number of
commonly used approximate methods, in addition to high force and mass
resolution simulations.  In this appendix we give the technical details
for the various mock catalog schemes, other than QPM, presented in this paper.

\subsection{Log-normal model} \label{app:lognormal}

A popular method for creating mock catalogs is the log-normal model of
\citet{ColJon91}.  This model has been used to generate mock catalogs
for analyses of the 2dF \citep[e.g.,][]{Col05},
SDSS \citep[most recently,][]{Per10,Rei10},
WiggleZ \citep{Bla11},
and 6dF \citep{Beu11,Beu12} surveys, among others.

\begin{figure}
\begin{center}
\resizebox{3.25in}{!}{\includegraphics{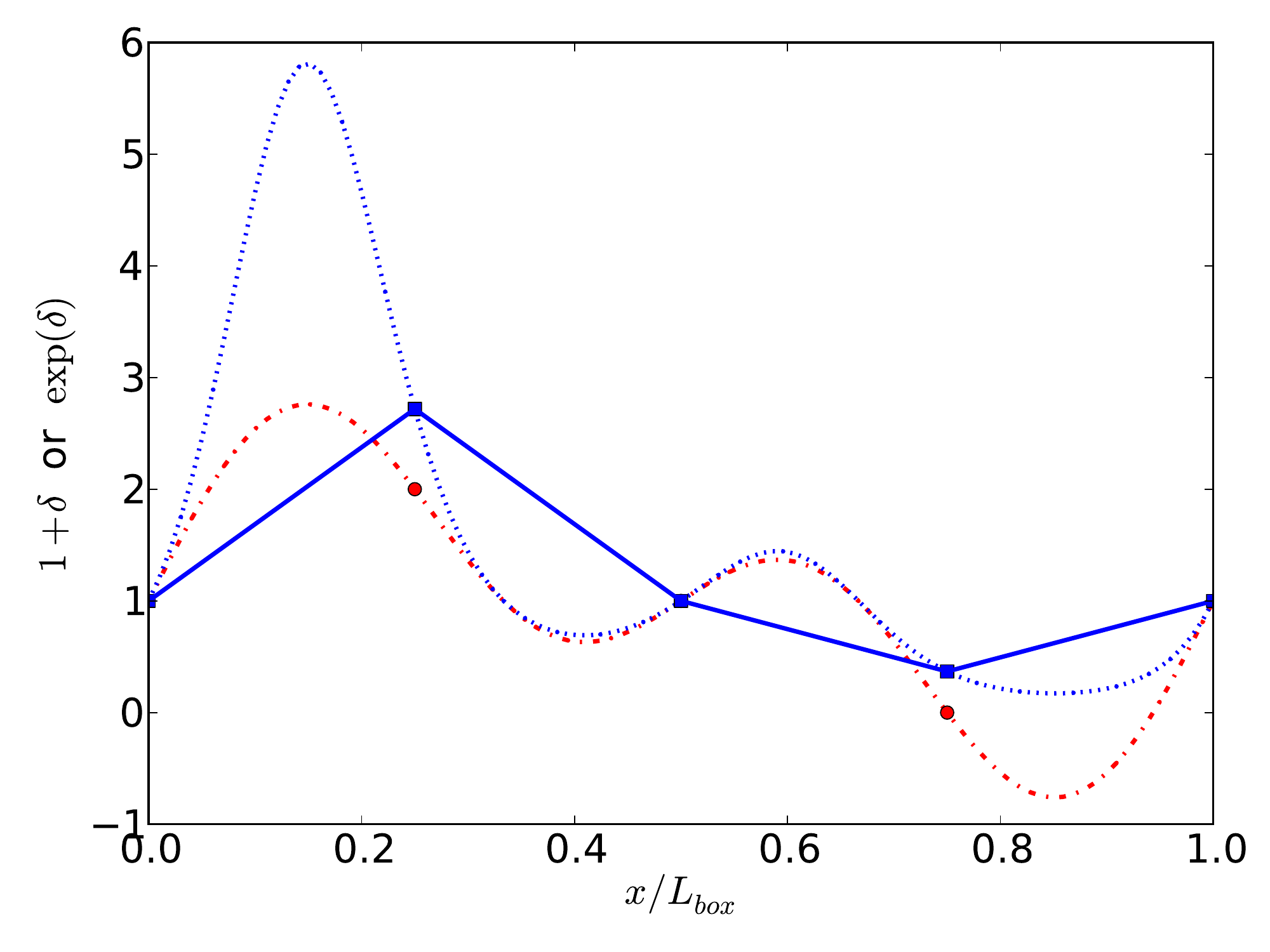}}
\end{center}
\caption{Sampling the LGN field: the $x$-axis shows the periodic
interval $[0,1)$ sampled by $N_g=4$ points.
The red dot-dashed line shows $1+\delta$ for a density constrast which is
the sum of the fundamental and Nyquist mode, each with unit amplitude:
$\delta(x)=\sin(2\pi x)+\sin(4\pi x)$.  The red circles show the value of
the field at the grid points.  The blue dotted lines show $\exp(\delta)$ for
the full field and the blue squares show the field sampled at the $N_g=4$ grid
points.  Note the grid Nyquist samples the input and the field is correct
at each grid point, but the the solid blue line (which shows the linear
interpolation of $\exp(\delta)$ on the grid) is significantly different from
the blue dotted line and the statistics of points generated from that
density field will differ significantly from those for points generated from
the full field.  To capture the information either a finer grid or a different
interpolation is needed.}
\label{fig:ln_example}
\end{figure}

In this model a Gaussian density field is generated with a correlation
function $\xi_G$ and then the field is exponentiated to obtain a log-normal
field which is then sampled (with probability proportional to density) to
produce a set of particles.
In order to obtain any desired $\xi_{LN}$ the correlation function
for the Gaussian field should obey $\xi_G=\ln(1+\xi_{LN})$.
The density distribution of the resulting particle set is log-normal, while
its two point function matches the desired function by construction.  These
particles thus behave in their low order statistics very similarly to the
halos in N-body simulations or galaxies in observational surveys.

Often the log-normal model is used to produce fields in real space, ignoring
redshift-space distortions.  There are two ways of introducing redshift-space
distortions into the model.  First, one can follow the reasoning in
\citet{ColJon91} who point out that the log-normal model can be thought of as
a kinematical model in which the velocity field is assumed to remain always
linear.  In this case the continuity equation can be solved for the density,
which is an exponential in the velocity divergence, $\theta$.  In linear theory
$\theta$ is simply proportional to $\delta$, which is assumed to obey Gaussian
statistics.  Thus we can draw our velocities from the Gaussian field,
$\delta_G$, which is exponentiated to give $\delta_{LN}$ by treating it as
(proportional to) the velocity divergence and generating
$\mathbf{v}(\mathbf{k})\propto (\mathbf{k}/k^2)\delta_G(\mathbf{k})$.
Such an approach has been used in \citet{KitGalFer12} for example.

A second method assumes a model for the effect of redshift-space distortions
on the correlation function (or power spectrum) of the Gaussian field and
simply generates a random realization of a Gaussian field from the anisotropic
2-point function.  This approach is followed in \citet{Beu12} using the
Kaiser approximation \citep{Kai87} for the power spectrum.
We follow the first approach rather than the second, because there are
some situations where it is useful to have the velocity information explicitly.

Our implementation of a log-normal code takes as input a target, real-space
correlation function and a value of $\beta\equiv f(\Omega)/b$.
We generate this correlation function using CLPT, as described in \citet{CLPT}.
The power spectrum of the Gaussian field is computed semi-analytically by
integrating $\ln(1+\xi_{LN})$ against $\sin(kr)/(kr)$ and then a Gaussian
random field is generated from this power spectrum in the usual way.
A large number of points are thrown randomly throughout the volume, and kept
with probability proportional to the density.  We interpolate the density
field to the trial position using ``cloud-in-cell'' interpolation
\citep[e.g.][]{HocEas80}.

We find that the Gaussian power spectrum damps strongly at
$k\simeq 1\,h\,{\rm Mpc}^{-1}$ but that if we wish to sample the density
field with tracers so as to obtain a point set then using grids with a finer
spacing provides better agreement between the 2-point function of the
generated point set and the target than a coarser grid
(for futher discussion see Fig.~\ref{fig:ln_example}).
If the analysis is done directly on the gridded field, or if the interpolation
scheme doesn't ``clip the peaks'', then the coarser grid is adequate.
As the velocity field is generated from the Gaussian velocity divergence
the field in redshift-space approaches the prediction of \citet{Kai87} on
large scales.

\subsection{LPT model} \label{app:2lpt}

The use of second order Lagrangian perturbation theory
\citep[LPT;][]{Buc89,Mou91,Hiv95,Sco98}
to generate density distributions in which halos are placed is the basis
of the PThalos algorithm \citep{PThalos}.
A variant of this approach, in which the halos in the LPT density field
are found using the friends-of-friends algorithm \citep[FoF;][]{DEFW},
was introduced in \citet{Man13}.

We follow the method outlined in \citet{Man13}.  From an initial
power spectrum we generate displacements using LPT.  We then displace
particles from an initially Cartesian grid along these displacement vectors
to form a particle-based realization of the density field.
We identify halos using the FoF algorithm and assign masses to the halos
by abundance matching to the mass function of an N-body simulation
(or a fit to it).

There are several choices which must be made in this algorithm
\citep[see also][for related discussion]{Ney12,Lec13}.
First we need to decide whether to damp the input power spectrum at high $k$.
While there are good arguments for such a truncation in some contexts, we
found that using an undamped spectrum is better in our implementation of this
algorithm.  This is because we displace particles from a Cartesian grid, and
a smoothed spectrum distorts the grid on large scales but leaves vestiges of
it clearly visible in the particle distribution on small scales.  When
coupled with a FoF halo finder working with a spherical linking criterion
this introduces undesirable artifacts.
A natural choice is to set the grid resolution to match the mean
inter-particle spacing.
If the particle loading and grid are too coarse we lose both mass resolution
and linear resolution.  Increasing the resolution of the grid too much can
also cause problems however, as at high resolution LPT fails\footnote{This
is particularly true using a Fourier-based estimator such as we employ.}
and leads to artifacts like halos in voids
\citep{Buc94,Bou95,SahSha96,Ney12,Lec13}.
A good compromise is a grid spacing and mean inter-particle spacing of
$\mathcal{O}(1\,h^{-1}{\rm Mpc})$ \citep[see also][]{Ney12}.
We use a $3000^3$ grid for the $2.75\,h^{-1}$Gpc boxes and a $1500^3$ grid
for the $1.5\,h^{-1}$Gpc boxes, with an equal number of particles as grid
points.

Having generated an initial density/particle field we need to partition the
mass into halos.  This step must necessarily be {\it ad hoc\/} because the
LPT method, like any perturbative scheme, generates the wrong small-scale
power and lacks information on halo formation.
These complexities are worse for hierarchical models such as cold dark matter
\citep[as compared with e.g.~hot dark matter;][]{MelShaWei94,MonTheTaf02}.
We follow \citet{Man13} and use the FoF algorithm.
This algorithm has a single parameter, the linking length, and produces a
unique partition of the particles into groups (of multiplicity 1 particle or
higher).  For a model such as CDM it is natural to express the linking length
in terms of the mean inter-particle separation, with typical numbers being
$0.1-0.2$ \citep{DEFW,LacCol94}.
In our situation the appropriate linking length to use (and whether it should
scale only with the mean inter-particle separation) is less clear.
We follow \citet{Man13} and empirically adjust the linking length to
match the clustering strength of a constant number density sample of halos
seen in our high-resolution N-body simulation.
Specifically, for the halo catalog produced at any linking length we rank
order the halos by particle number and choose those whose number densities
lie within a narrow range determined from the abundance of halos in the
high resolution simulation.  For halos spanning an octave in mass centered
on lg$M=12.5$, 13 and 13.5 ($h^{-1}M_\odot$) we find number densities of
$-3.24<{\rm lg}\,n<-2.95$, $-3.82<{\rm lg}\,n<-3.45$, $-4.09<{\rm lg}\,n<-4.54$,
($h^{-3}{\rm Mpc}^3$) in the $2.75\,h^{-1}$Gpc simulation.
As the linking length is increased the large-scale bias of each sample,
as determined by the amplitude of the halo-mass cross-spectrum at low $k$,
decreases.  We were unable to find a single linking length which matched
the large-scale bias for all 3 mass/abundance bins, so we chose the linking
length which best matched the $10^{13}\,h^{-1}M_\odot$ sample.  This linking
length was the same as found by \citet{Man13}, $b=0.36$.


\end{document}